\documentclass[12pt]{iopart}

\usepackage{iopams}  
\usepackage{graphicx}
\usepackage{amssymb}
\usepackage{color}
\usepackage{psfrag}

\newcommand{\bra}[2] {\mbox{}_{#2}\langle #1 |}
\newcommand{\ket}[2] {| #1 \rangle_{#2}}
\newcommand{\braket}[2] {\langle #1 | #2 \rangle}

\begin{document}

\title[Optimised generation of heralded Fock states using parametric down conversion]{Optimised generation of heralded Fock states using parametric down conversion}

\author{Agata M.~Bra{\'n}czyk$^{1,2}$, T. C. Ralph$^1$, Wolfram Helwig$^2$, Christine Silberhorn$^2$}

\address{$^1$Center for Quantum Computer Technology, Department of Physics, The University of Queensland,
QLD 4072, Australia\\$^2$Max-Planck-Institute for the Science of Light, G\"
unther-Scharowsky-Str. 1/Building 24, 91058 Erlangen, Germany }
\date{24 June 2008}

\ead{abranczyk@gmail.com}
\begin{abstract}
The generation of heralded pure Fock states via spontaneous parametric down conversion (PDC) relies on perfect photon-number correlations in the output modes. Correlations in any other degree of freedom, however, degrade the purity of the heralded state. In this paper, we investigate spectral entanglement between the two output modes of a periodically poled waveguide. With the intent of generating heralded 1- and 2-photon Fock states, we expand the output state of the PDC to second order in photon number. We explore the effects of spectral filtering and inefficient detection, of the heralding mode, on the count rate, $g^{(2)}$ and purity of the heralded state, as well as the fidelity between the resulting state and an ideal Fock state. We find that filtering can decrease spectral correlations, however, at the expense of the count rate and increased photon-number mixedness in the heralded output state. As a physical example, we model a type II PP-KTP waveguide pumped by lasers at wavelengths of $400$ nm, $788$ nm and $1.93~\mu$m. The latter two allow the fulfillment of extended phase matching conditions in an attempt to eliminate spectral correlations in the PDC output state without the use of filtering, however, we find that even in these cases, some filtering is needed to achieve states of very high purity. 

\end{abstract}

\vspace{2pc}
\maketitle

\section{Introduction}
Pure photon Fock states, in particular, single-photon states, are useful for many quantum-optical applications, including quantum information processing, quantum computing and quantum cryptography \cite{Kok2007}. Some novel uses for photon number states include the generation of other non-Gaussian states, such as Schr\"odinger kitten states \cite{Ourjoumtsev2006,Ourjoumtsev2007}. 

A common method for creating single-photon states makes use of spontaneous parametric down conversion (SPDC), where a CW or pulsed laser is used to pump a nonlinear crystal. Due to the nonlinear properties of such a crystal, photons in the pump laser are spontaneously down converted into pairs of lower energy photons. Momentum conservation ensures that the photon wave packets are well localised with respect to each other. The detection of a single photon in one spatial mode (idler) heralds the presence of another single photon in the other spatial mode (signal). In practice, however, given high enough pump power, the presence of higher order photon-number terms in the output state can lead to a photon-number mixed state in the signal mode, when inefficient detectors mistake two (or more) photons for one. 

Energy conservation ensures that the frequencies of the downconverted photons always sum to the pump frequency. For CW pumped PDC, these correlations cannot be avoided, but pulsed pump light allows this constraint to be weakened.  Strong spectral correlations are another potential source of mixedness -- the signal state is projected into a spectrally mixed state when a frequency-insensitive detector heralds a single photon in the signal mode. In the context of single-mode versus multi-mode descriptions \cite{Tualle-Brouri2009,Rohde2007}  this property can  also be interpreted as projecting the single photon onto different distinguishable broadband spectral modes \cite{URen2005}. In recent years, there has been a growing effort in engineering pulsed SPDC sources to produce photons uncorrelated in frequency, i.e. those with a separable joint spectral amplitude (JSA). Some examples include manipulating the crystal length, material, bandwidth and central frequency \cite{Kim2002, Grice2001, Walton2004, URen2006, Corona2007, URen2005, URen2007, Kuzucu2008, Mosley2008, Garay-Palmett2007} as well as filtering the pump field, prior to down-conversion, using an optical cavity \cite{Raymer2005}. Another promising technique produces a source of counter-propagating photons with a separable JSA \cite{Christ2009a}. 

Typical theoretical analyses of multimode effects in PDC truncate the output state to first order in photon number. In this paper, we will extend our analysis to include second-order photon number contributions. We will compare the more humble method of pre-detection filtering, of which the main shortcoming is the loss in the production rate, with a more experimentally challenging approach where the JSA is made as separable as possible by satisfying various phase matching conditions. We find that even if these conditions are fulfilled, some level of filtering is still required in order to achieve states with very high purities. For both methods, we characterise the effects of higher order photon number contributions, on the generation of single-photon states. In addition, we consider the production of two-photon Fock states, conditional on the detection of two photons in the idler mode. 

As a practical example, we consider a periodically poled KTP (PP-KTP) wave guide -- this eliminates any spatial correlations -- pumped by ultrafast optical pulses to ensure that the pairs exhibit tight localisations in time. 

This paper is structured as follows. In section \ref{sec:PDC}, we introduce spectral notation for photon states as well as a theoretical model of spectral effects in SPDC. In section \ref{sec:filter} we introduce spectral filtering. In section \ref{sec:1Fock} we present analytical results for the probability of detecting a single photon in the heralding detector, the heralded output state, its $g^{(2)}$ and purity, as well as the maximum fidelity between the heralded output state and an ideal pure state. In section \ref{sec:2Fock}, we present similar results for the generation of two-photon Fock states. In sections \ref{sec:phys1} and \ref{sec:phys2}, we illustrate these results using realistic parameters. In section \ref{sec:conc} we discuss our results. 

Finally, a note on nomenclature. In our theoretical calculations, we have a preference for using frequency (as opposed to wavelength) due to its direct relationship to energy conservation in SPDC, however, we have made an attempt to also present our results in nanometers (nm), for readers who are accustomed to ``thinking in wavelengths''. All frequencies quoted in this paper are \emph{angular frequencies} in units of $\mathrm{s}^{-1}$. When describing Gaussian filters and beam profiles, we will specify the central frequency and Gaussian standard deviation in $\mathrm{s}^{-1}$ as well as the central wavelength and the FWHM in nm.

\section{Parametric Down Conversion}\label{sec:PDC}
A single-mode PDC can be modeled in the interaction picture, where the evolution of the state vector is given by \cite{Walls1984}
\begin{eqnarray}
\ket{\psi(t)}{}=\exp(\kappa t (\hat{a}_{i}^{\dagger}\hat{a}_{s}^{\dagger}-\hat{a}_{i}\hat{a}_{s}))\ket{\psi(t_0)}{}\,.
\end{eqnarray}
This gives the output of a PDC \footnote{Note that a PDC state of one spatio-spectral mode is exactly equivalent to a two-mode squeezed beam.}, in the number basis, as
\begin{eqnarray}
\ket{\psi_{\mathrm{out}}}{}=\frac{1}{\mathrm{cosh}(\kappa t)}\sum_{n=0}^{\infty}\mathrm{tanh}(\kappa t)^n\ket{n}{i}\ket{n}{s}\,,
\end{eqnarray}
where $\kappa$ is the effective nonlinearity and is a function of the pump power and the properties of the nonlinear crystal and $t$ is the interaction time. The output state is correlated in photon number. 

To take the spectral properties of the system into consideration, we define the creation operator for a photon with a spectral distribution $\psi(\omega)$ as \cite{Rohde2007,Rohde2005}:
\begin{eqnarray}
\hat{A}^{\dagger}_{\psi}=\int d\omega \psi(\omega) \hat{a}^{\dagger}(\omega)\,.
\end{eqnarray} 
$\hat{A}^{\dagger}_{\psi}$ satisfies all the standard bosonic commutation relations, such as $[\hat{A}_{\psi_k},\hat{A}^{\dagger}_{\psi_{k'}}]=\delta_{k,k'}$, where $\psi_k(\omega)$ and $\psi_{k'}(\omega)$ are orthogonal spectral functions, i.e. $\int \psi_k(\omega) \psi_{k'}(\omega) d\omega\nolinebreak=\nolinebreak\delta_{k,k'}$. An $n$-photon state can be written as
\begin{eqnarray}
\ket{n;\psi}{}=\frac{1}{\sqrt{n!}}(\hat{A}^{\dagger}_{\psi})^n\ket{0}{}\,.
\end{eqnarray}
We emphasise the distinction between the states $\hat{A}^{\dagger}_{\psi_k}\hat{A}^{\dagger}_{\psi_k}=\sqrt{2}\ket{2;\psi_k}{}$ and $\hat{A}^{\dagger}_{\psi_k}\hat{A}^{\dagger}_{\psi_{k'}}\nolinebreak=\nolinebreak\ket{1;\psi_k}{}\ket{1;\psi_{k'}}{}$ for $k\neq k'$, where the former is a two-photon Fock state and the latter consists of two single-photon Fock states.

For type-II down conversion, where the pump is non-delpleting, i.e. classical, we can take the multimode Hamiltonian to be \cite{Grice1997}
\begin{eqnarray}
H(t)=\int_V d^3 r \chi^{(2)} E^{(+)}_p(\textbf{r},t)\hat{E}^{(-)}_i(\textbf{r},t)\hat{E}^{(-)}_s(\textbf{r},t)+\mathrm{H.c.}\,,
\end{eqnarray}
where $V$ is the spatial mode volume in the waveguide and $\hat{E}_j(\textbf{r},t)=\hat{E}^{(+)}_j(\textbf{r},t)+\hat{E}^{(-)}_j(\textbf{r},t)$ are the three interacting fields with $j=p,i,s$ denoting the pump, idler and signal modes respectively. 
\begin{eqnarray}
E^{(+)}_p(z,t)&=&A_p\int d\omega_p\alpha(\omega_p)\mathrm{e}^{i[k_p(\omega_p)z+\omega_pt]}\\
\hat{E}^{(-)}_j(z,t)&=&\int d\omega_j A(\omega_j)\hat{a}^{\dagger}(\omega_j)\mathrm{e}^{-i[k_j(\omega_j)z+\omega_jt]}\,,
\end{eqnarray}
where $\alpha(\omega_p)=\exp(-(\omega_p-\mu_p)^2/2\sigma_p)$ is the pump envelope function and we have restricted the spatial integral to be over only one dimension, ie. $z$, and $j=i,s$. This Hamiltonian does not commute with itself at different times and therefore, the evolution of the state vector should be taken to be
\begin{eqnarray}\label{eq:unitary} 
\ket{\psi(t)}{}=U(t)\ket{\psi(t_0)}{}=\mathcal{T}\mathrm{e}^{{-}\frac{\imath}{\hbar}\int _{t_0}^{t} dt'H(t')}\ket{\psi(t_0)}{}\,,
\end{eqnarray}
where $\mathcal{T}$ is the time-ordering operator. $U(t)$ can be expanded into what is known as the Dyson series. This is very challenging, and as an approximation, we will drop the time-ordering operator and expand $U(t)$ as a Taylor series. In the scenario presented in this paper, we do not expect this type of approximation to be problematic, however, we acknowledge that problems may arise when considering input states of a quantum nature, as has been investigated by Leung \emph{et al.} \cite{Leung2008}. A description of multiple pair creation in degenerate SPDC has also been analysed in the Heisenberg picture by Wasilewki \emph{et. al} \cite{Wasilewski2006} and Mauerer \cite{Mauerer2009}.

We will assume that $A(\omega_j)$ is slowly varying over the frequencies of interest and therefore we can bring it outside of the integral. We can now write
\begin{eqnarray}\nonumber
\int _{t_0}^{t} dt'H(t')&=&A\int _{{-}\infty}^{\infty} dt'\int_{{-}L/2}^{L/2}dz\int d\omega_i  d\omega_s d\omega_p e^{-i[k_i(\omega_i)+k_s(\omega_s)-k_p(\omega_p)]z}\\&&\times e^{i[\omega_i+\omega_s-\omega_p]t}\alpha(\omega_p)\hat{a}_i^{\dagger}(\omega_i)\hat{a}_s^{\dagger}(\omega_s)+\mathrm{H.c.}\,,
\end{eqnarray}
where $L$ is the length of the crystal and $A=\chi^{(2)}A_pA(\omega_i)A(\omega_s)$. For a pulsed laser, we can assume that the pump field, and therefore the interaction Hamiltonian, is zero before $t_0$ and after $t$. Therefore we can extend the limits of the integration over time to $-\infty$ and $\infty$ \cite{Boyd2003}. Performing the time integral yields $2\pi\delta(\omega_i+\omega_s-\omega_p)$ which then allows the $\omega_p$ integral to be evaluated, giving
\begin{eqnarray}\nonumber
\int _{t_0}^{t} dt'H(t')&=&-2\pi A\int_{{-}L/2}^{L/2}dz\int d\omega_i d\omega_s\hat{a}_i^{\dagger}(\omega_i)\hat{a}_s^{\dagger}(\omega_s)\\
&&\times\alpha(\omega_i+\omega_s)e^{{-}i[k_i(\omega_i)+k_s(\omega_s)-k_p(\omega_i+\omega_s)]z}+\mathrm{H.c.}
\end{eqnarray}
Evaluating the integral over $z$ yields
\begin{eqnarray}\label{eq:ham}
\int _{t_0}^{t} dt'H(t')&=&2\pi AL\int d\omega_i d\omega_s\hat{a}_i^{\dagger}(\omega_i)\hat{a}_s^{\dagger}(\omega_s)\alpha(\omega_i+\omega_s)\Phi(\omega_i,\omega_s)+\mathrm{H.c.}\,,
\end{eqnarray}
where $\Phi(\omega_i,\omega_s)$ is the phase-matching function
\begin{eqnarray}
\Phi(\omega_i,\omega_s)=\mathrm{sinc}\Big( \frac{ L\Delta k}{2}\Big)\,,\label{eq:PMF}
\end{eqnarray}
where $\mathrm{sinc}(x)=\sin(x)/x$ and $ \Delta k = k_i(\omega_i)+k_s(\omega_s)-k_p(\omega_i+\omega_s)$.  For a periodically poled waveguide of periodicity $\Lambda$, $ \Delta k = k_i(\omega_i)+k_s(\omega_s)-k_p(\omega_i+\omega_s)+2\pi/\Lambda$ \cite{Eckstein2006}. Note that by picking the spatial integration to be centered around $z=0$, it is possible to eliminate a phase term which would normally be present in equation (\ref{eq:ham}). For the experiment, this corresponds to pre-chirping the pump pulse with an adapted phase progression.   

Following Grice and Walmsley \cite{Grice1997}, we Taylor expand the phase mismatch to first order such that $\Delta k\approx \Delta k^{(0)}+k_s'\nu_s+k_i'\nu_i-k_p'\nu_p$ where $\nu_j=\omega_j-\mu_j$, $k_j'=\partial k_j(\omega)/\partial \omega |_{\omega=\mu_j}$ and  $\mu_j$ is the center frequency of a photon in mode $j$. We set $\mu_i = \mu_s = \mu$ and $\mu_p=2\mu$.  We can achieve perfect phase-matching by picking $\Lambda$ such that $ \Delta k^{(0)} = k_s(\mu_s)+k_i(\mu_i)-k_p(\mu_p)=2\pi/\Lambda$ and therefore $\Delta k\approx k_s'\nu_s+k_i'\nu_i-k_p'\nu_p$.

To consider contributions from the 2-photon components of the down-converted state, we take the Taylor series expansion of the unitary evolution operator in equation (\ref{eq:unitary}) to second order (disregarding the time-ordering operator):
\begin{eqnarray}
U(t)\approx1+\frac{1}{i\hbar}\int _{t_0}^{t} dt_1H(t_1)+\frac{1}{2(i\hbar)^2}\int _{t_0}^{t} dt_2H(t_2)\int _{t_0}^{t} dt_3H(t_3)\,.
\end{eqnarray}
This gives the downconverted state
\begin{eqnarray}\nonumber
\ket{\psi_{\mathrm{PDC}}}{}&=&N\Big\{\big(1+\chi^2\big)\ket{0}{}+\chi\int\int d\omega_i d\omega_s  f(\omega_i,\omega_s) \hat{a}_i^{\dagger}(\omega_i)\hat{a}_s^{\dagger}(\omega_s)\ket{0}{}\\\nonumber
&&+\frac{\chi^2}{2}\int\int d\omega_i d\omega_s  f(\omega_i,\omega_s) \hat{a}_i^{\dagger}(\omega_i)\hat{a}_s^{\dagger}(\omega_s)\\
&&\times\int\int d\omega'_i d\omega'_s  f(\omega'_i,\omega'_s) \hat{a}_i^{\dagger}(\omega'_i)\hat{a}_s^{\dagger}(\omega'_s)\ket{0}{}\Big\}\,,
\end{eqnarray}
where $\chi=2\pi AL/i\hbar$ and $N$ is defined in equation (\ref{eq:N}). The joint spectral amplitude (JSA) is given by
\begin{eqnarray}\label{eq:JSA}
f(\omega_i,\omega_s)=N_f\alpha(\omega_i+\omega_s)\Phi(\omega_i,\omega_s)\,,
\end{eqnarray}
 where the normalisation parameter $N_f$ is chosen such that $\int d\omega_i d\omega_s  |f(\omega_i,\omega_s)|^2=1$.
   
Any well-behaved complex function can always be decomposed 
in terms of a discrete basis of orthonormal functions (a well known example is the basis of Hermite 
functions). This is known as the Schmidt decomposition.
\begin{eqnarray}
f(\omega_s,\omega_i)= \sum_k b_k\xi_k(\omega_i)\zeta_k(\omega_s)\,,
\end{eqnarray}
where the Schmidt modes $\xi_k(\omega_i)$ and $\zeta_k(\omega_s)$ are normalised and may be complex and the Schmidt coefficients $b_k$ are real and $\sum_k|b_k|^2=1$, if $f(\omega_s,\omega_i)$ is normalised. It is useful to write the downconverted state in terms of the Schmidt decomposition (refer to Table \ref{tab:modes} for creation operator definitions).
\begin{eqnarray}\nonumber
\ket{\psi_{\mathrm{PDC}}}{}&=&N\Big\{\big(1+\chi^2\big)\ket{0}{}+\chi\sum_k b_k \hat{A}^{\dagger}_{i_{\xi_k}}\hat{A}^{\dagger}_{s_{\zeta_k}}\ket{0}{}\\\label{eq:schmidt_total}
&&+\frac{\chi^2}{2}\sum_{k,k'}b_{k}b_{k'}\hat{A}^{\dagger}_{i_{\xi_{k}}}\hat{A}^{\dagger}_{i_{\xi_{k'}}}\hat{A}^{\dagger}_{s_{\zeta_{k}}}\hat{A}^{\dagger}_{s_{\zeta_{k'}}}\ket{0}{}\Big\}\,,
\end{eqnarray}
where $\xi_{k}(\omega_i)$ are the Schmidt modes for the idler state and $\zeta_{k}(\omega_s)$ are the Schmidt modes for the signal state and 
\begin{eqnarray}\label{eq:N}
N=\Big\{\big|1+\chi^2\big|^2+|\chi|^2+|\chi|^4\Big(\sum_{\scriptsize\begin{array}{c} k,k' \\ k{<}k'\end{array}\normalsize}\big|b_{k} b_{k'}\big|^2+\sum_{k}\big|b_{k}\big|^4\Big)\Big\}^{-1/2}\,.
\end{eqnarray}
Notice, in the four-photon term of equation (\ref{eq:schmidt_total}), when two photons are created in the same spectral mode (i.e. $k=k'$) there will be a factor of $\sqrt{2}$ in front of each two-photon Fock state, increasing the probability of down conversion into such a state. This can be understood due to stimulation effects in the PDC process itself. Equation (\ref{eq:schmidt_total}) can also be written as

\begin{eqnarray}\nonumber
\ket{\psi_{\mathrm{PDC}}}{}&=&N\Big\{\big(1+\chi^2\big)\ket{0}{}+\chi\sum_k b_k \ket{1;\xi_k}{i}\ket{1;\zeta_k}{s}\\\nonumber
&&+\chi^2\Big(\sum_{k}b_{k}^2 \ket{2;\xi_{k}}{i}\ket{2;\zeta_{k}}{s}\\\label{eq:schmidt_total2}
&&+\sum_{\scriptsize\begin{array}{c} k,k' \\ k{<}k'\end{array}\normalsize}b_{k} b_{k'} \ket{1;\xi_{k}}{i}\ket{1;\xi_{k'}}{i}\ket{1;\zeta_{k}}{s}\ket{1;\zeta_{k'}}{s}\Big)\Big\}\,.
\end{eqnarray}
We can characterise the spectral entanglement of the JSA by using the entropy of entanglement \cite{Bennett1996}. The entropy of entanglement can be defined, for the bi-partite state
\begin{eqnarray}
\ket{\Psi}{}=\sum_k b_k \ket{1;\xi_k}{}\ket{1;\zeta_k}{}
\end{eqnarray}
in terms of the Schmidt values:
\begin{eqnarray}
E(\ket{\Psi}{})=-\sum_kb_k^2\log_2(b_k^2)\,.
\end{eqnarray}
The entropy of entanglement is valid only for pure bipartite states and, when defined in terms of the Schmidt decomposition, can not be applied to the entire output state in equation (\ref{eq:schmidt_total2}). However, we can apply it to the two-photon term to get some information about the spectral entanglement arising only from the JSA. The entropy of entanglement ranges from zero for a product state to $\log_2N$ for a maximally entangled state of two $N$-state particles, which in our case corresponds to a state containing $N$ orthogonal spectral modes. In the limit of a maximally entangled JSA, the entropy of entanglement would be $\infty$. 

\begin{table}[t]
\centering
\begin{tabular}{c c c l}
\hline
\hline
Creation  & Spectral & State&  Description\\
Operator & Mode &&\\
\hline
\hline
$\hat{A}^{\dagger}_{i_{\zeta_k}}$ & $\zeta_k(\omega_i)$ &$\ket{1;\zeta_k}{i}$& initial idler Schmidt modes\\
$\hat{A}^{\dagger}_{s_{\xi_k}}$ & $\xi_k(\omega_s)$ & $\ket{1;\xi_k}{s}$&initial signal Schmidt modes\\
$\hat{C}^{\dagger}_{T\zeta_{k} }$ & $T(\omega_i)\zeta_k(\omega_i)$&$\ket{1;T\zeta_k}{i}$ & filtered idler modes\\
$\hat{D}^{\dagger}_{R\zeta_{k} }$ & $R(\omega_i)\zeta_k(\omega_i)$&$\ket{1;R\zeta_k}{i}$  & reflected filtered idler modes\\
$\hat{C}^{\dagger}_{\phi_{j} }$ & $\phi_j(\omega_i)$& $\ket{1;\phi_j}{i}$& orthog. filtered idler modes\\
$\hat{D}^{\dagger}_{\varphi_{j} }$ & $\varphi_j(\omega_i)$  &$\ket{1;\varphi_j}{i}$& orthog. reflected filtered idler modes\\
$\hat{A}^{\dagger}_{i_{\tau_k}} $& $\tau_m(\omega_s)$ &$\ket{1;\tau_m}{s}$& diag. single-photon signal modes \\
\hline
\hline
\end{tabular}
\caption{Summary of multi-mode creation operators, spectral modes and states.}
\label{tab:modes}
\end{table}

\section{Spectral Filtering}\label{sec:filter}

A spectral filter can be modeled as a frequency dependent beam-splitter:
\begin{eqnarray}
\hat{a}^{\dagger}(\omega)\rightarrow\tilde{T}(\omega)\hat{c}^{\dagger}(\omega)+\tilde{R}(\omega)\hat{d}^{\dagger}(\omega)\,,
\end{eqnarray}
where $|\tilde{T}(\omega)|^2$ and $|\tilde{R}(\omega)|^2$ are the transmitted and reflected probabilities and  $|\tilde{T}(\omega)|^2+|\tilde{R}(\omega)|^2=1$.  In addition to the filter, we consider an inefficient detector which we model by a beam splitter of reflectivity $1-\eta$, followed by a perfect detector (refer to figure \ref{fig:PDC_schematic}). If the reflected mode of the filter and the reflected mode of the beamsplitter are to be traced out, the filter-beamsplitter combination can be modeled by a filter with the following transformation:
\begin{eqnarray}\label{eq:spectral_filt}
\hat{a}^{\dagger}(\omega)\rightarrow T(\omega)\hat{c}^{\dagger}(\omega)+R(\omega)\hat{d}^{\dagger}(\omega)\,,
\end{eqnarray}
where $T(\omega)=\tilde{T}(\omega)\sqrt{\eta}$ and $R(\omega)=\sqrt{1-|\tilde{T}(\omega)|^2\eta}$. In terms of the mode functions $\zeta_k$, this can be written as 
\begin{eqnarray}\label{eq:filter}
\hat{A}^{\dagger}_{\zeta_{k}}&\rightarrow&T_{\zeta_{k}} \hat{C}^{\dagger}_{T\zeta_{k} }+R_{\zeta_{k}}\hat{D}^{\dagger}_{R\zeta_{k}}
\end{eqnarray}
where we have defined
\begin{eqnarray}\label{eq:T_zeta_k}
T_{\zeta_{k}}&=&\sqrt{\int d\omega |T(\omega)\zeta_{k}(\omega)|^2}\\
R_{\zeta_{k}}&=&\sqrt{\int d\omega |R(\omega)\zeta_{k}(\omega)|^2}\\
\hat{C}^{\dagger}_{T\zeta_{k} }&=&\frac{1}{T_{\zeta_{k}}}\int d\omega T(\omega)\zeta_{k}(\omega)\hat{c}^{\dagger}(\omega)\\\label{eq:D_T_zeta_k}
\hat{D}^{\dagger}_{R\zeta_{k} }&=&\frac{1}{R_{\zeta_{k}}}\int d\omega R(\omega)\zeta_{k}(\omega)\hat{d}^{\dagger}(\omega)\,.
\end{eqnarray}
The definitions in equations (\ref{eq:T_zeta_k})-(\ref{eq:D_T_zeta_k}) ensure that the creation operators for the filtered modes satisfy the commutation relations $[\hat{C}_{T\zeta_{k} },\hat{C}^{\dagger}_{T\zeta_{k} }]=1$ and $[\hat{D}_{R\zeta_{k} },\hat{D}^{\dagger}_{R\zeta_{k} }]=1$. However, the filtered functions $T\zeta_{k}(\omega_s)$ no longer define proper modes because the functions  $\zeta_k(\omega_s)T(\omega_s)$ are, in general, not orthogonal to the functions $\zeta_{k'}(\omega_s)T(\omega_s)$ for $k\neq k'$ and therefore need to be orthogonalised (eg. using the Gram-Schmidt procedure) such that:
 \begin{eqnarray}
 T_{\zeta_{k}}\hat{C}^{\dagger}_{T\zeta_{k} }\ket{0}{}&=&\sum_{j} u_{kj}   \hat{C}^{\dagger}_{ \phi_j}\ket{0}{} \\
R_{\zeta_{k}}\hat{D}^{\dagger}_{T\zeta_{k} }\ket{0}{}&=&\sum_{j} v_{kj}   \hat{D}^{\dagger}_{ \varphi_j}\ket{0}{}\,,
 \end{eqnarray}
 where $\phi_j(\omega_i)$ are now the new modes defining the idler state and $\varphi_j(\omega_i)$ are the reflected modes that will be traced out, and 
 \begin{eqnarray}\label{eq:ukj}
 u_{kj}&=&\int d\omega \phi_j(\omega)^*\zeta_{k}(\omega)T(\omega)=T_{\zeta_{k}}\braket{1;\phi_j}{1;T\zeta_{k}}\\\label{eq:vkj}
 v_{kj}&=&\int d\omega \varphi_j(\omega)^*\zeta_{k}(\omega)R(\omega)=R_{\zeta_{k}}\braket{1;\varphi_j}{1;R\zeta_{k}}\,.
 \end{eqnarray}
The filter relationship in equation (\ref{eq:filter}) can now be written as follows:
 \begin{eqnarray}\label{eq:filter2}
\hat{A}^{\dagger}_{{\zeta_k}}&\rightarrow& \sum_{j} \Big(u_{kj}   \hat{C}^{\dagger}_{ \phi_j}+v_{kj}   \hat{D}^{\dagger}_{ \varphi_j}\Big)\,.
\end{eqnarray}
Because the filter has been modeled as a frequency dependent beamsplitter, it will have similar properties to a beamsplitter. One property, that we will evoke throughout this paper, is the tendency to allow one photon, from an incident two-photon state, to pass through the filter while rejecting the other. Detection of a filtered two-photon state may then result in, even, a perfect detector mistaking it for a one-photon state. 

\section{Generating single-photon Fock states}\label{sec:1Fock}

Detection of a single photon in the idler mode heralds the presence of a single photon in the signal mode. In this section, we consider a hypothetical perfect detector, an inefficient detector and a Gaussian spectral filter placed in front of the lossy detector. 

\subsection{Case 1: Perfect Detection in the triggering idler mode}\label{sec:det1}

\begin{figure}[t!]
 \begin{center}
  \vspace{-.2cm}
 \includegraphics[width=14cm]{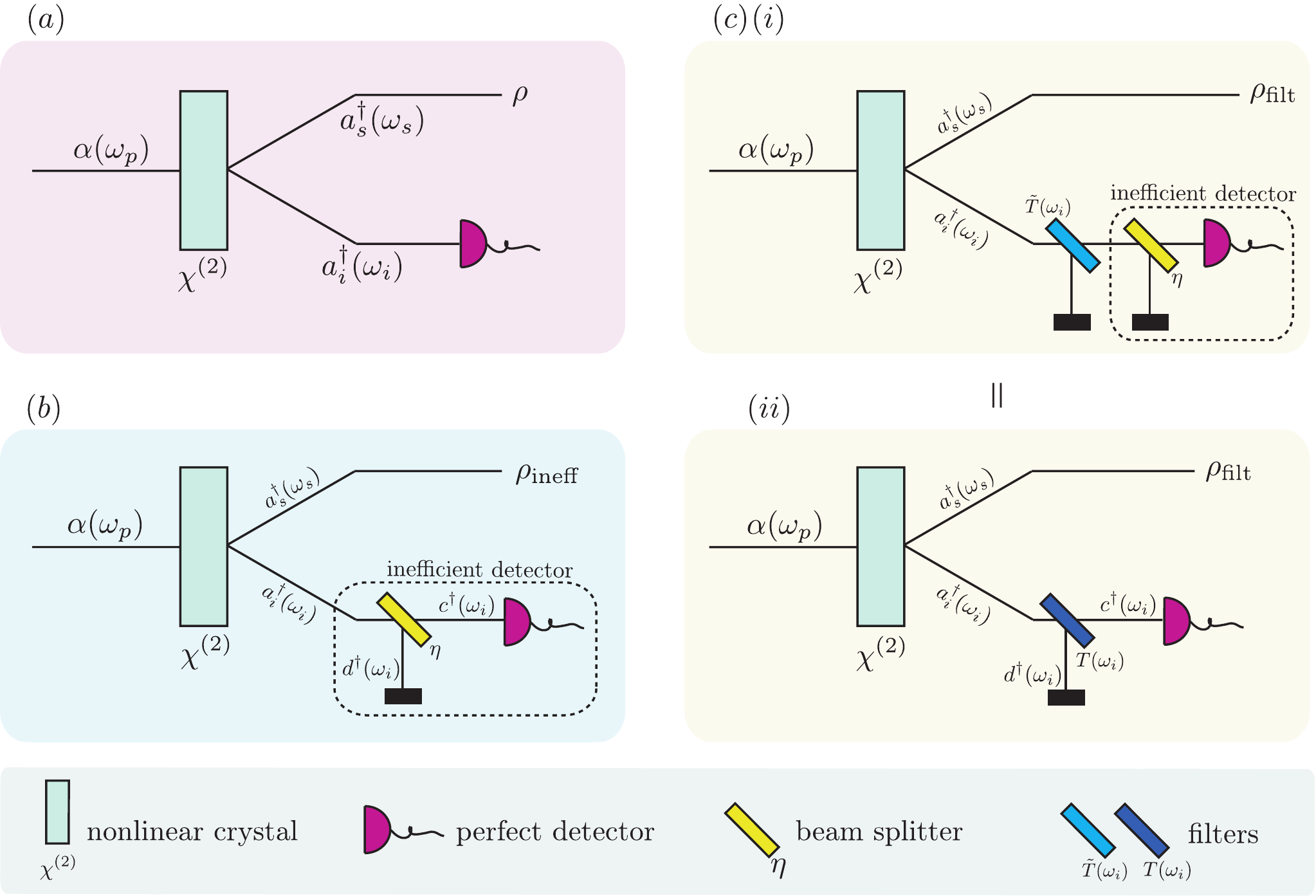} 
   \vspace{-.5cm}
\end{center}
  \caption{Schematic diagrams of SPDC setup for: (a) perfect detection in the triggering mode (see sections \ref{sec:det1} and \ref{sec:det2}); (b) inefficient detection in the triggering mode (see sections \ref{sec:IneffDet1} and \ref{sec:IneffDet2}); and (c)(i) filtering the triggering mode prior to detection with inefficient detector and (ii) the equivalent setup where the filter and beamsplitter have been combined into one filter (see sections \ref{sec:FiltDet1} and \ref{sec:FiltDet2}).} 
  \label{fig:PDC_schematic} 
\end{figure}

The POVM for a detector that perfectly distinguishes photon number, but gains no information about the frequency of the photon can be written as follows:
\begin{eqnarray}\label{eq:POVM_1_a}
\Pi_1 = \int d\omega \hat{a}^{\dagger}(\omega)\ket{0}{}\bra{0}{}\hat{a}(\omega)= \sum_{j}\ket{1;\xi_j}{}\bra{1;\xi_j}{}\,.
\end{eqnarray}
We can interpret this as: the detection of a single photon $\hat{a}^{\dagger}(\omega)\ket{0}{}$, however due to the lack of spectral knowledge, $\omega$ must be integrated over; or alternatively, the detection of a single photon in the spectral mode $\ket{1;\xi_j}{}$, however due to the lack of knowledge about which mode it was in, it is necessary to sum over $j$.  Since photon detection is destructive, the detected mode must be traced out. The probability of detecting a single photon in the idler mode using a perfect single-photon detector (refer to figure \ref{fig:PDC_schematic}(a)), is
\begin{eqnarray}\label{eq:P_det_1}
p_{\mathrm{1}}=\bra{\Psi_{\mathrm{PDC}}}{}\Pi_1\ket{\Psi_{\mathrm{PDC}}}{}=|N|^2|\chi|^2\,.
\end{eqnarray}
 Given a single-photon detection in the idler mode, the heralded signal state is
\begin{eqnarray}
\rho_{{\mathrm{1}}}=\frac{1}{p_{\mathrm{1}}}\mathrm{Tr}_{i}\big[\Pi_1\ket{\Psi_{\mathrm{PDC}}}{}\bra{\Psi_{\mathrm{PDC}}}{}\Pi_1\big]=\sum_k |b_k|^2 \ket{1;\xi_k}{s}\bra{1;\xi_k}{}\,.
\end{eqnarray}
The $g^{(2)}$ of the signal state, which we define as
\begin{eqnarray}\label{eq:g2}
g^{(2)}&=&\frac{\sum_{j,j'}\langle\hat{A}_{{\xi_j}}^{\dagger}\hat{A}_{{\xi_j}'}^{\dagger}\hat{A}_{{\xi_j}}\hat{A}_{{\xi_j}'}\rangle}{\Big(\sum_{j}\langle\hat{A}_{{\xi_j}}^{\dagger}\hat{A}_{{\xi_j}}\rangle\Big)^2}\,.
\end{eqnarray}
is $g^{(2)}=0$. This reveals that there is only one photon in the signal mode, but not how pure it is. The purity of the heralded state is
\begin{eqnarray}
P_1=\mathrm{Tr}[\rho_{{1}}^2]=|N_{1}|^4\sum_k |b_k|^4\,.
\end{eqnarray} 
For a state which only contains single photons, i.e. one that is heralded by a perfect detector, the purity is equivalent to the Hong-Ou-Mandel visibility \cite{Hong1987}. 

Without loss of generality, we order the Schmidt coefficients in decreasing order from $k=0$. Therefore, the pure single photon state with the highest overlap with the projected state will be the photon mode corresponding to the highest Schmidt coefficent, $b_0$, and hence the maximum fidelity with a single photon Fock state is
\begin{eqnarray}\label{eq:F_det_1}
F_1=\max_jF(\rho_{{\mathrm{1}}}, \ket{1;\xi_j}{})=\bra{1;\xi_0}{}\rho_{{\mathrm{1}}}\ket{1;\xi_0}{}=|b_0|^2\,.
\end{eqnarray}
The fidelity does not depend on $\chi$. Increasing the strength of the nonlinearity only has an effect on how often the detector registers a \emph{click}, however, once that happens, the signal mode is always projected into the same state.

\subsection{Case 2: Inefficient Detection in the triggering idler mode}\label{sec:IneffDet1}
An inefficient detector can be modeled by the transformation in equation (\ref{eq:spectral_filt}) where $T(\omega)=\sqrt{\eta}$ and $R(\omega)=\sqrt{1-\eta}$, followed by a perfect detector (refer to figure \ref{fig:PDC_schematic}(b)). After the beam splitter, the joint signal-idler state is
\begin{eqnarray}
\rho_{\mathrm{ineff}}=\mathrm{Tr}_{\hat{D}}[\ket{\Psi_{\mathrm{ineff}}}{}\bra{\Psi_{\mathrm{ineff}}}{}]\,,
\end{eqnarray}
where
\begin{eqnarray}\nonumber
\ket{\Psi_{\mathrm{ineff}}}{}&=&N\Big\{\big(1+\chi^2\big)\ket{0}{}+\chi\sum_k b_k \big(\sqrt{\eta}\hat{A}^{\dagger}_{i_{\xi_k}}\hat{C}^{\dagger}_{{\zeta_k}}+\sqrt{1-\eta}\hat{A}^{\dagger}_{i_{\xi_k}}\hat{D}^{\dagger}_{{\zeta_k}}\big)\ket{0}{}\\\nonumber
&&+\frac{\chi^2}{2}\Big(\sum_{k,k'}b_{k}b_{k'}\hat{A}^{\dagger}_{i_{\xi_{k}}}\hat{A}^{\dagger}_{i_{\xi_{k'}}}\Big(\eta\hat{C}^{\dagger}_{{\zeta_{k}}}\hat{C}^{\dagger}_{{\zeta_{k'}}}+(1-\eta)\hat{D}^{\dagger}_{{\zeta_{k}}}\hat{D}^{\dagger}_{{\zeta_{k'}}}\\
&&+\sqrt{\eta}\sqrt{1-\eta}(\hat{C}^{\dagger}_{{\zeta_{k'}}}\hat{D}^{\dagger}_{{\zeta_{k}}}+\hat{C}^{\dagger}_{{\zeta_{k}}}\hat{D}^{\dagger}_{{\zeta_{k'}}})\Big)\ket{0}{}\Big)\Big\}\,.
\end{eqnarray}
The probability of detecting a single photon in the idler mode, using an inefficient single-photon detector, is
\begin{eqnarray}
p_{\mathrm{1,ineff}}&=&\mathrm{Tr}[\Pi_1\rho_{\mathrm{ineff}}]\\\label{eq:P_ineff_det_1}
&=&|N|^2|\chi|^2\eta\Big\{1+2|\chi|^2(1-\eta)\Big(\sum_{\scriptsize\begin{array}{c} k,k' \\ k{<}k'\end{array}\normalsize}|b_{k}b_{k'}|^2+\sum_{k}|b_{k}|^4\Big)\Big\}\,.
\end{eqnarray}
Given a single-photon detection in the idler mode, the heralded signal state is
\begin{eqnarray}\label{eq:ineffDet1}
\rho_{{\mathrm{1,ineff}}}&=&\frac{1}{p_{\mathrm{1,ineff}}}\mathrm{Tr}_{\hat{C}}\big[\Pi_1\rho_{\mathrm{ineff}}\Pi_1\big]\\\nonumber
&=&|N_{\mathrm{1,ineff}}|^2\Big\{\sum_k |b_k|^2\ket{1;\xi_k}{i}\bra{1;\xi_k}{}\\\nonumber
&&+2|\chi|^2(1-\eta)\Big(\sum_{k}|b_{k}|^4\ket{2;\xi_{k}}{i}\bra{2;\xi_{k}}{}\\
&&+\sum_{\scriptsize\begin{array}{c} k,k' \\ k{<}k'\end{array}\normalsize}|b_{k}b_{k'}|^2\ket{1;\xi_{k}}{i}\ket{1;\xi_{k'}}{i}\bra{1;\xi_{k}}{}\bra{1;\xi_{k'}}{i}\Big)\Big\}\,,
\end{eqnarray}
where $N_{\mathrm{1,ineff}}=N\chi\sqrt{\eta}/\sqrt{p_{\mathrm{1,ineff}}}$. The $g^{(2)}$ for this state, defined in equation (\ref{eq:g2}), is
\begin{eqnarray}\label{eq:g2_ineff}
g^{(2)}=\frac{\gamma_i}{|N_{\mathrm{1,ineff}}|^{2}(1+\gamma_i)^2}\,,
\end{eqnarray}
where
\begin{eqnarray}
\gamma_i=4|\chi|^2(1-\eta)\Big(\sum_{\scriptsize\begin{array}{c} k,k' \\ k{<}k'\end{array}\normalsize}|b_{k}b_{k'}|^2+\sum_{k'}|b_{k}|^4\Big)\,.
\end{eqnarray}
The purity of the signal state is
\begin{eqnarray}
P_{{\mathrm{1,ineff}}}&=&\mathrm{Tr}[\rho_{{\mathrm{1,ineff}}}^2]\\\nonumber
&=&|N_{\mathrm{det,1,ineff}}|^4\Big\{\sum_k |b_k|^4+4|\chi|^4(1-\eta)^2\\\label{eq:Pur_ineff_det_1}
&&\times\Big(\sum_{\scriptsize\begin{array}{c} k,k' \\ k{<}k'\end{array}\normalsize}|b_{k}b_{k'}|^4+\sum_{k}|b_{k}|^8\Big)\Big\}\,.
 \end{eqnarray} 
The maximum fidelity, between the heralded state and a pure Fock state $\ket{1;\xi_j}{}$, is
\begin{eqnarray}\label{eq:F_ineff_det_1}
F_{\mathrm{1,ineff}}=\max_jF(\ket{1;\xi_{j}}{},\rho_{{\mathrm{1,ineff}}})=\bra{1;\xi_{0}}{}\rho_{{\mathrm{1,ineff}}}\ket{1;\xi_{0}}{}=|N_{\mathrm{1,ineff}}|^2|b_0|^2
\end{eqnarray}

\subsection{Case 3: Filtering the idler state}\label{sec:FiltDet1}
We now introduce a filter in the idler mode as shown in figure \ref{fig:PDC_schematic}(c). Applying a filter, as defined in equation (\ref{eq:filter}), to the signal mode, gives the filtered state

\begin{eqnarray}\label{eq:filtDet1}
\rho_{\mathrm{filt}}=\textrm{Tr}_{\hat{D}}[\ket{\Psi_{\mathrm{filt}}}{}\bra{\Psi_{\mathrm{filt}}}{}]\,,
\end{eqnarray}
where
\begin{eqnarray}\nonumber
\ket{\Psi_{\mathrm{filt}}}{}&=&N\Big\{\big(1+\chi^2\big)\ket{0}{}+\chi\sum_{k,j} b_k \hat{A}^{\dagger}_{i_{\xi_k}}\Big(u_{kj}   \hat{C}^{\dagger}_{ \phi_j}+v_{kj}   \hat{D}^{\dagger}_{ \varphi_j}\Big)\ket{0}{}\\\nonumber
&&+\frac{\chi^2}{2}\sum_{k,k',j,j'}b_{k}b_{k'}\hat{A}^{\dagger}_{i_{\xi_{k}}}\hat{A}^{\dagger}_{i_{\xi_{k'}}}\Big(u_{kj}u_{k'j'}   \hat{C}^{\dagger}_{ \phi_{j}}\hat{C}^{\dagger}_{ \phi_{j'}}+v_{kj}u_{k'j'}   \hat{D}^{\dagger}_{ \varphi_{j}}\hat{C}^{\dagger}_{ \phi_{j'}}\\
&&+u_{kj}v_{k'j'}    \hat{C}^{\dagger}_{ \phi_{j}} \hat{D}^{\dagger}_{ \varphi_{j'}}+v_{kj} v_{k'j'}   \hat{D}^{\dagger}_{ \varphi_{j}} \hat{D}^{\dagger}_{ \varphi_{j'}}\Big)\ket{0}{}\Big\}\,,
\end{eqnarray}
and $u_{kj}$ and $v_{kj}$ are defined as per equations (\ref{eq:ukj}) and (\ref{eq:vkj}).  The probability of detecting a single photon in the filtered idler mode is
\begin{eqnarray}
p_{\mathrm{1,filt}}&=&\mathrm{Tr}[\Pi_1\rho_{\mathrm{filt}}]\\\nonumber
&=&|N|^2|\chi|^2\Big\{\sum_{k} |b_k|^2T^2_{\zeta_{k}}+|\chi|^2\Big\{\sum_k2|b_{k}|^4T^2_{\zeta_{k}}R^2_{\zeta_{k}}\\\label{eq:P_filt_det_1}
&&+\sum_{\scriptsize\begin{array}{c} k,k' \\ k{<}k'\end{array}\normalsize}|b_{k}b_{k'}|^2\Big(T^2_{\zeta_{k}}R^2_{\zeta_{k'}}+ T^2_{\zeta_{k'}}R^2_{\zeta_{k}}+\mathrm{T}_{kk'}\mathrm{R}_{k'k} +\mathrm{T}_{k'k} \mathrm{R}_{kk'}\Big)\Big\}\,,
\end{eqnarray}
where
\begin{eqnarray}
\mathrm{T}_{kk'}&=&T_{\zeta_k}T^*_{\zeta_{k'}} \braket{1; T{\zeta_{k'}}}{1; T{\zeta_k}}=\sum_ju_{kj}u^*_{k'j}\\
\mathrm{R}_{kk'}&=&R_{\zeta_{k}}R^*_{\zeta_{k'}} \braket{1; R{\zeta_{k'}}}{1; R{\zeta_{k}}}=\sum_jv_{kj}v^*_{k'j}\,.
\end{eqnarray}
Given a single-photon detection in the idler mode, the heralded signal state is
\begin{eqnarray}\label{eq:rho_det_1_ineff}
\rho_{{\mathrm{1,filt}}}&=&\frac{1}{p_{\mathrm{1,filt}}}\mathrm{Tr}_{C}\big[\Pi_1\rho_{\mathrm{filt}}\Pi_1\big]\\\nonumber
&=&|N_{\mathrm{1,filt}}|^2\Big\{\sum_{k,\tilde{k}}b_kb^*_{\tilde{k}} \mathrm{T}_{k\tilde{k}} \ket{1;\xi_k}{}\bra{1;\xi_{\tilde{k}}}{}\\\label{eq:rho_filt_det_1}
&&+|\chi|^2\Big( \sum_{k,k',\tilde{k},\tilde{k}' }b_{k}b_{k'}b^*_{\tilde{k}}b^*_{\tilde{k}'}\mathrm{T}_{k\tilde{k}'}\mathrm{R}_{k'\tilde{k}}\hat{A}^{\dagger}_{\xi_k}\hat{A}^{\dagger}_{\xi_{k'}}\ket{0}{}\bra{0}{}\hat{A}_{\xi_{\tilde{k}}}\hat{A}_{\xi_{\tilde{k}'}}\Big)\,,
\end{eqnarray}
where $N_{\mathrm{1,filt}}=N\chi/\sqrt{p_{\mathrm{1,filt}}}$.  Note that filtering the idler mode also changes the mode structure of the heralded signal state.  The $g^{(2)}$ for this state, defined in equation (\ref{eq:g2}), is
\begin{eqnarray}\label{eq:g2_filt}
g^{(2)}=\frac{\gamma_f}{|N_{\mathrm{1,filt}}|^{2}(\sum_{k}|b_k|^2|T_{\zeta_k}|^2+\gamma_f)^2}\,,
\end{eqnarray}
where
\begin{eqnarray}\nonumber
\gamma_f&=&2|\chi|^2\Big\{2\sum_{k}|b_{k}|^4\mathrm{R}_{kk}\mathrm{T}_{kk}\\
&&+\sum_{\scriptsize\begin{array}{c} k,k' \\ k<k'\end{array}\normalsize }|b_{k}b_{k'}|^2(\mathrm{R}_{kk}\mathrm{T}_{k'k'}+\mathrm{R}_{k'k}\mathrm{T}_{kk'}+\mathrm{R}_{kk'}\mathrm{T}_{k'k}+\mathrm{R}_{k'k'}\mathrm{T}_{kk})\Big\}\,.
\end{eqnarray}
The purity of the heralded state is
\begin{eqnarray}
P_{\mathrm{1,filt}}&=&\mathrm{Tr}[\rho_{{\mathrm{1,filt}}}^2]\\\nonumber
&=&|N_{\mathrm{1,filt}}|^4\Big\{\sum_{k,\tilde{k}} |b_kb_{\tilde{k}}|^2| \mathrm{T}_{k\tilde{k}}|^2 +|\chi|^4\Big\{\sum_{\tilde{k}}2|b_{\tilde{k}}|^{4}\\\nonumber
&&\times\Big(\hspace{-2mm}\sum_{\scriptsize\begin{array}{c} k,k' \\ k<k'\end{array}\normalsize }|b_{k}b_{k'}|^2|\mathrm{R}_{k\tilde{k}}\mathrm{T}_{k'\tilde{k}}+\mathrm{R}_{k'\tilde{k}}\mathrm{T}_{k\tilde{k}}|^2+\sum_{k}2|b_{k}|^4|\mathrm{R}_{k\tilde{k}}\mathrm{T}_{k\tilde{k}}|^2\Big)\\\nonumber
&&+\hspace{-2mm}\sum_{\scriptsize\begin{array}{c} \tilde{k},\tilde{k}' \\ \tilde{k}<\tilde{k}'\end{array}\normalsize }\hspace{-2mm}|b^*_{\tilde{k}}b^*_{\tilde{k}'}|^2\Big(\sum_{k}2|b_{k}|^4|\mathrm{R}_{k\tilde{k}}\mathrm{T}_{k\tilde{k}'}{+}\mathrm{R}_{k\tilde{k}'}\mathrm{T}_{k\tilde{k}}|^2\\\label{eq:Pur_filt_det_1}
&&+\hspace{-2mm}\sum_{\scriptsize\begin{array}{c} k,k' \\ k<k'\end{array}\normalsize }\hspace{-2mm}|b_{k}b_{k'}|^2|\mathrm{R}_{k\tilde{k}}\mathrm{T}_{k'\tilde{k}'}{+}\mathrm{R}_{k'\tilde{k}}\mathrm{T}_{k\tilde{k}'}{+}\mathrm{R}_{k\tilde{k}'}\mathrm{T}_{k'\tilde{k}}{+}\mathrm{R}_{k'\tilde{k}'}\mathrm{T}_{k\tilde{k}}|^2\Big)\Big\}\Big\}\,.
\end{eqnarray}
The density matrix in equation (\ref{eq:rho_filt_det_1}) is not diagonal in the $\ket{1;\xi_k}{}$ basis, but this can be easily achieved, for the part of the state which is relevant for calculating the fidelity with a single-photon Fock state, giving
\begin{eqnarray}\label{eq:rho_det_1_ineff}
\rho_{{\mathrm{1,filt,part}}}&=&|N_{\mathrm{1,filt}}|^2\sum_{k,\tilde{k}}b_kb^*_{\tilde{k}} \mathrm{T}_{k\tilde{k}} \ket{1;\xi_k}{}\bra{1;\xi_{\tilde{k}}}{}=\sum_md_m\ket{1;\tau_m}{}\bra{1;\tau_m}{}\,,
\end{eqnarray}
where
\begin{eqnarray}\label{eq:tau}
\ket{1;\tau_m}{}&=&\sum_kc_{mk}\ket{1;\xi_k}{}\,.
\end{eqnarray}
and $\tau_m(\omega_s)$ are the new orthogonal modes defining the signal state. The maximum fidelity, between the heralded state and a pure single-photon state $\ket{1;\tau_l}{}$, is
\begin{eqnarray}\label{eq:F_filt_det_1}
F_{\mathrm{1,filt}}=\max_l F(\ket{1;\tau_l}{},\rho_{{\mathrm{1,filt}}})=\max_l\bra{1;\tau_{l}}{}\rho_{{\mathrm{1,filt,part}}}\ket{1;\tau_{l}}{}=\max_m d_m\,.
\end{eqnarray}
In an experiment, the spectral distribution $\tau_m$ should be chosen in any interferometric experiment to optimise for the best performance of the heralded single photons. 

In the extreme case where $\tilde{T}(\omega)=\delta(\omega-\mu)$, i.e. the filter picks out a single frequency $\mu$,  the fidelity tends to unity and the signal state tends to the pure state
\begin{eqnarray}
\ket{\Psi_{\mathrm{1,filt},\delta}}{}=N_{\mathrm{1,filt}}\sum_{k}b_k \zeta_k(\mu)\ket{1;\xi_{k}}{}=\sqrt{d_m}\ket{1;\tau_{m}}{}\,,
\end{eqnarray}
as $\chi\rightarrow 0$. This implies that it is possible to obtain arbitrarily pure single photon states, with the use of spectral filtering and by ensuring the nonlinearity strength is low.

\section{Generating 2-photon Fock states}\label{sec:2Fock}

In addition to creating single-photon states, it is becoming increasingly desirable to create higher photon-number Fock states. In this section we will investigate the effects of detector efficiency, and filtering of the idler mode, on the generation of two-photon Fock states in the signal mode conditional on the detection of heralded two-photon states in the idler mode. 

\subsection{Case 1: Perfect Detection in the triggering idler mode}\label{sec:det2}

The projector for detecting two photons in any spectral modes $\zeta_{j}$ and $\zeta_{j'}$ will be separated into two parts: the part which detects two photons in orthogonal modes and the part which detects two photons in the same mode:
\begin{eqnarray}
\Pi_2&=&\sum_{j}\ket{2;\zeta_{j}}{}\bra{2;\zeta_{j}}{}+\sum_{\scriptsize\begin{array}{c} j,j'\\ j{<}j'\end{array}\normalsize}\ket{1;\zeta_{j}}{}\ket{1;\zeta_{j}}{}\bra{1;\zeta_{j}}{}\bra{1;\zeta_{j}}{}\,.
\end{eqnarray}  
Refer to the schematic in figure \ref{fig:PDC_schematic}(a). The probability of detecting two photons in the idler mode, with a frequency insensitive detector, is
\begin{eqnarray}\label{eq:P_det_2}
p_{\mathrm{2}}=\bra{\Psi_{\mathrm{PDC}}}{}\Pi_2\ket{\Psi_{\mathrm{PDC}}}{}=|N|^2|\chi|^4\Big(\sum_{\scriptsize\begin{array}{c} k,k'\\ k{<}k'\end{array}\normalsize}|b_{k} b_{k'} |^2 +\sum_{k}|b_{k}|^4\Big)\,.
\end{eqnarray}
Given a two-photon detection in the idler mode, the heralded state in the signal mode is
\begin{eqnarray}
\rho_{{\mathrm{2}}}&=&\frac{1}{p_{\mathrm{2}}}\mathrm{Tr}_{i}\big[\Pi_2\ket{\Psi_{\mathrm{PDC}}}{}\bra{\Psi_{\mathrm{PDC}}}{}\Pi_2\big]\\\nonumber
&=&|N_{\mathrm{2}}|^2\Big\{\sum_{k} |b_{k}|^4 \ket{2;\xi_{k}}{s}\bra{2;\xi_{k}}{}\\
&&+\sum_{\scriptsize\begin{array}{c} k,k'\\ k{<}k'\end{array}\normalsize} |b_{k}b_{k'}|^2 \ket{1;\xi_{k}}{s}\ket{1;\xi_{k'}}{s}\bra{1;\xi_{k}}{}\bra{1;\xi_{k'}}{}\Big\}\\
&=&\frac{|N_{\mathrm{2}}|^2}{2}\sum_{k}|b_{k}|^2  \hat{A}^{\dagger}_{s_{\xi_k}}\ket{0}{}\bra{0}{}\hat{A}_{s_{\xi_k}}\otimes\sum_{k'}|b_{k'}|^2\hat{A}^{\dagger}_{s_{\xi_{k'}}}\ket{0}{}\bra{0}{}\hat{A}_{s_{\xi_{k'}}}\,,
\end{eqnarray}
where $N_{\mathrm{2}}=N\chi^2/\sqrt{p_{\mathrm{2}}}$. The purity is
\begin{eqnarray}
P_{\mathrm{2}}=\textrm{Tr}[\rho_{\mathrm{2}}^2]=|N_{\mathrm{2}}|^4\Big\{\sum_{\scriptsize\begin{array}{c} k,k'\\ k{<}k'\end{array}\normalsize} |b_{k}b_{k'}|^4 +\sum_{k} |b_{k}|^8 \Big\}\,.
\end{eqnarray}
The $g^{(2)}$ for this state is $g^{(2)}=1/2$. It is interesting to note that the $g^{(2)}$ does not depend on the purity of the two-photon state. It will always remain at the value of $1/2$ regardless of whether the two-photon state is in a Fock state or in some other form. This reflects the fact that $g^{(2)}$ is only sensitive to the photon number, but not the modal properties of the state.

The maximum fidelity between the heralded state and an ideal two-photon Fock state $\ket{2;\xi_j}{}$ is
\begin{eqnarray}\label{eq:fid_two_phot}
F_{\mathrm{2}}=\max_j F(\rho_{{\mathrm{2}}}, \ket{2;\xi_j}{})=\bra{2;\xi_0}{}\rho_{{\mathrm{2}}}\ket{2;\xi_0}{}=|N_{\mathrm{2}}|^2|b_0|^4\,.
\end{eqnarray}

\subsection{Case 2: Inefficient Detection in the triggering idler mode}\label{sec:IneffDet2}

The probability of detecting two photons in the idler mode, with an inefficient frequency insensitive detector (refer to figure \ref{fig:PDC_schematic}(b)), is
\begin{eqnarray}\label{eq:P_ineff_det_2}
p_{\mathrm{2,ineff}}=\mathrm{Tr}[\Pi_2\rho_{\mathrm{ineff}}]=|N|^2|\chi|^4\eta^2\Big(\sum_{\scriptsize\begin{array}{c} k,k'\\ k{<}k'\end{array}\normalsize}|b_{k} b_{k'} |^2 +\sum_{k}|b_{k}|^4\Big)\,.
\end{eqnarray}
where $\rho_{\mathrm{ineff}}$ is defined in equation (\ref{eq:ineffDet1}).  Because our analysis only extends to second order in photon-number, the expressions for the fidelity and purity will be the same as they were in section \ref{sec:det2}, where a perfect detector was used. If we included higher order terms, we would expect the fidelity and purity to vary as a function of $\chi$ and $\eta$ in a similar fashion to the single-photon case in section \ref{sec:1Fock}.

\subsection{Case 3: Filtering of the idler state}\label{sec:FiltDet2}

We now introduce a filter in the idler mode as shown in figure \ref{fig:PDC_schematic}(c). After filtering the state, the probability of detecting two photons in the idler mode, with an inefficient frequency insensitive detector, is
\begin{eqnarray}
p_{\mathrm{2,filt}}&=&\mathrm{Tr}[\Pi_2\rho_{\mathrm{filt}}]\\\nonumber
&=&|N|^2|\chi|^4\Big(\sum_{k'}|b_{k}|^4||T_{\zeta_{k}}|^4\\\label{eq:P_filt_det_2}
&&+\sum_{\scriptsize\begin{array}{c} k,k'\\ k{<}k'\end{array}\normalsize}|b_{k}b_{k'}|^2(|T_{\zeta_{k}}|^2|T_{\zeta_{k'}}|^2 +\mathrm{T}_{kk'}\mathrm{T}_{k'k})\Big)\,,
\end{eqnarray}
where $\rho_{\mathrm{filt}}$ is defined in equation (\ref{eq:filtDet1}). Given a two-photon detection in the idler mode, the heralded state in the signal mode is\newpage
\begin{eqnarray}
\rho_{{\mathrm{2,filt}}}&=&\frac{1}{p_{\mathrm{2}}}\mathrm{Tr}_{C}\big[\Pi_2\rho_{\mathrm{filt}}\Pi_2\big]\\\nonumber
\hspace{-0.3cm}&=&|N_{\mathrm{2,filt}}|^2\Big\{\sum_{k,\tilde{k}}b_{k}^2b_{\tilde{k}}^{*2}\mathrm{T}_{k\tilde{k}}^2\ket{2;\xi_k}{}\bra{2;\xi_{\tilde{k}}}{}\\\nonumber
\\\nonumber
&&+\hspace{-2mm}\sum_{{\scriptsize\begin{array}{c} k,k',\tilde{k}\\ k{<}k'\end{array}\normalsize}}\hspace{-0.3cm}\sqrt{2}b_{k}b_{k'}b_{\tilde{k}}^{*2}\mathrm{T}_{k\tilde{k}}\mathrm{T}_{k'\tilde{k}} \ket{1;\xi_k}{}\ket{1;\xi_{k'}}{}\bra{2;\xi_{\tilde{k}}}{}\\\label{eq:rho_filt_det_2}
&&+\hspace{-2mm}\sum_{\scriptsize\begin{array}{c} k, \tilde{k},\tilde{k}'\\ \tilde{k}{<}\tilde{k}'\end{array}\normalsize}\hspace{-0.3cm}\sqrt{2}b_{k}^2b^*_{\tilde{k}}b^*_{\tilde{k}'}\mathrm{T}_{k\tilde{k}}\mathrm{T}_{k\tilde{k}'} \ket{2;\xi_k}{} \bra{1;\xi_{\tilde{k}}}{}\bra{1;\xi_{\tilde{k}'}}{}\\\nonumber
&&+\hspace{-5mm}\sum_{{\scriptsize\begin{array}{c} k,k',\tilde{k},\tilde{k}'\\ k{<}k',\tilde{k}{<}\tilde{k}'\end{array}\normalsize}}\hspace{-0.3cm}b_{k}b_{k'}b^*_{\tilde{k}}b^*_{\tilde{k}'}(\mathrm{T}_{k\tilde{k}}\mathrm{T}_{k'\tilde{k}'}  +\mathrm{T}_{k'\tilde{k}}\mathrm{T}_{k\tilde{k}'})\ket{1;\xi_k}{}\ket{1;\xi_{k'}}{} \bra{1;\xi_{\tilde{k}}}{}\bra{1;\xi_{\tilde{k}'}}{}\Big\}\\\nonumber
&=&\frac{|N_{\mathrm{2,filt}}|^2}{2}\sum_{k,\tilde k}b_k b^*_{\tilde k}\textrm{T}_{k,\tilde{k}}\hat{A}^{\dagger}_{s_{\xi_k}}\ket{0}{}\bra{0}{}\hat{A}_{s_{\xi_{\tilde k}}}\\
&&\otimes\sum_{k',\tilde k'}b_{k'} b^*_{\tilde k'}\textrm{T}_{k',\tilde{k}'}\hat{A}^{\dagger}_{s_{\xi_{k'}}}\ket{0}{}\bra{0}{}\hat{A}_{s_{\xi_{\tilde k'}}}\,,
\end{eqnarray}
where $N_{\mathrm{2,filt}}=N\chi^2/\sqrt{p_{\mathrm{2,filt}}}$. Again $g^{(2)}=1/2$ as there are always two photons in the state. The purity of the heralded state is
\begin{eqnarray}\nonumber
P_{\mathrm{2,filt}}&=&\textrm{Tr}[\rho_{\mathrm{2,filt}}^2]=|N_{\mathrm{2,filt}}|^4\Big\{\sum_{\scriptsize\begin{array}{c} k,k'\\ k{<}k'\end{array}\normalsize}|b_{k}b_{k'}|^2\Big(\sum_{\tilde{k}}2|b_{\tilde{k}}|^{4}|\textrm{T}_{k,\tilde{k}}\textrm{T}_{k',\tilde{k}}|^2\\\nonumber
&&+\sum_{\scriptsize\begin{array}{c} \tilde{k},\tilde{k}'\\ \tilde{k}{<}\tilde{k}'\end{array}\normalsize}|b_{\tilde{k}}b_{\tilde{k}'}|^2|\textrm{T}_{k,\tilde{k}}\textrm{T}_{k',\tilde{k}'}  +\textrm{T}_{k',\tilde{k}}\textrm{T}_{k,\tilde{k}'}|^2 \Big)\\
&&+\sum_{k}|b_{k}|^4\Big(\sum_{\tilde{k}}|b_{\tilde{k}}|^{4}|\textrm{T}_{k,\tilde{k}}|^4+\sum_{\scriptsize\begin{array}{c} \tilde{k},\tilde{k}'\\ \tilde{k}{<}\tilde{k}'\end{array}\normalsize}2|b_{\tilde{k}}b_{\tilde{k}'}|^2|\textrm{T}_{k,\tilde{k}}\textrm{T}_{k,\tilde{k}'}|^2\Big)\Big\}\,.
\end{eqnarray}
The density matrix in equation (\ref{eq:rho_filt_det_2}) is not diagonal in the $\ket{1;\xi_k}{}$ basis. This can be easily achieved, giving
\begin{eqnarray}\label{eq:rho_det_1_ineff}
\rho_{{\mathrm{2,filt}}}&=&\frac{1}{2}\sum_{m,m'}d_md_{m'}\hat{A}^{\dagger}_{s_{\tau_m}}\hat{A}^{\dagger}_{s_{\tau_{m'}}}\ket{0}{}\bra{0}{}\hat{A}_{s_{\tau_m}}\hat{A}_{s_{\tau_{m'}}}\\
&=&\sum_{m}d_m^2\ket{2;\tau_m}{}\bra{2;\tau_{m}}{}\\
&&+\sum_{{\scriptsize\begin{array}{c} m,m'\\ m{<}m'\end{array}\normalsize}}d_md_{m'}\ket{1;\tau_m}{}\ket{1;\tau_{m'}}{}\bra{1;\tau_m}{}\bra{1;\tau_{m'}}{}\,,
\end{eqnarray}
where $\ket{1;\tau_m}{}$ is defined in equation (\ref{eq:tau}). The maximum fidelity, between the heralded state and a pure two-photon state with an optimised spectral distribution function $\ket{2;\tau_l}{}$, is
\begin{eqnarray}
F_{\mathrm{2,filt}}=\max_l F(\ket{2;\tau_l}{},\rho_{{\mathrm{2,filt}}})=\max_l\bra{2;\tau_{l}}{}\rho_{{\mathrm{2,filt}}}\ket{2;\tau_{l}}{}=\max_m d_m^2\,.
\end{eqnarray}
In the extreme case where $\tilde{T}(\omega)=\delta(\omega-\mu)$, i.e. the filter picks out a single frequency $\mu$,  the fidelity tends to unity and the signal state tends to the pure state
\begin{eqnarray}
\ket{\Psi_{\mathrm{2,filt},\delta}}{}&=&\frac{N_{\mathrm{2,filt}}}{2}\sum_{k}b_k \zeta_k(\mu)\hat{A}^{\dagger}_{s_{\xi_{k}}}\ket{0}{}\otimes\sum_{k'}b_{k'} \zeta_{k'}(\mu)\hat{A}^{\dagger}_{s_{\xi_{k'}}}\ket{0}{}\\
&=&d_m\ket{2;\tau_{m}}{}\,,
\end{eqnarray}
as $\chi\rightarrow 0$. This implies that it is possible to obtain arbitrarily pure two-photon states, with the use of spectral filtering and by ensuring the nonlinearity strength is low.

\section{Physical example I - correlated JSA}\label{sec:phys1}

\begin{figure}[t]
\begin{center}
  \includegraphics[width=16cm]{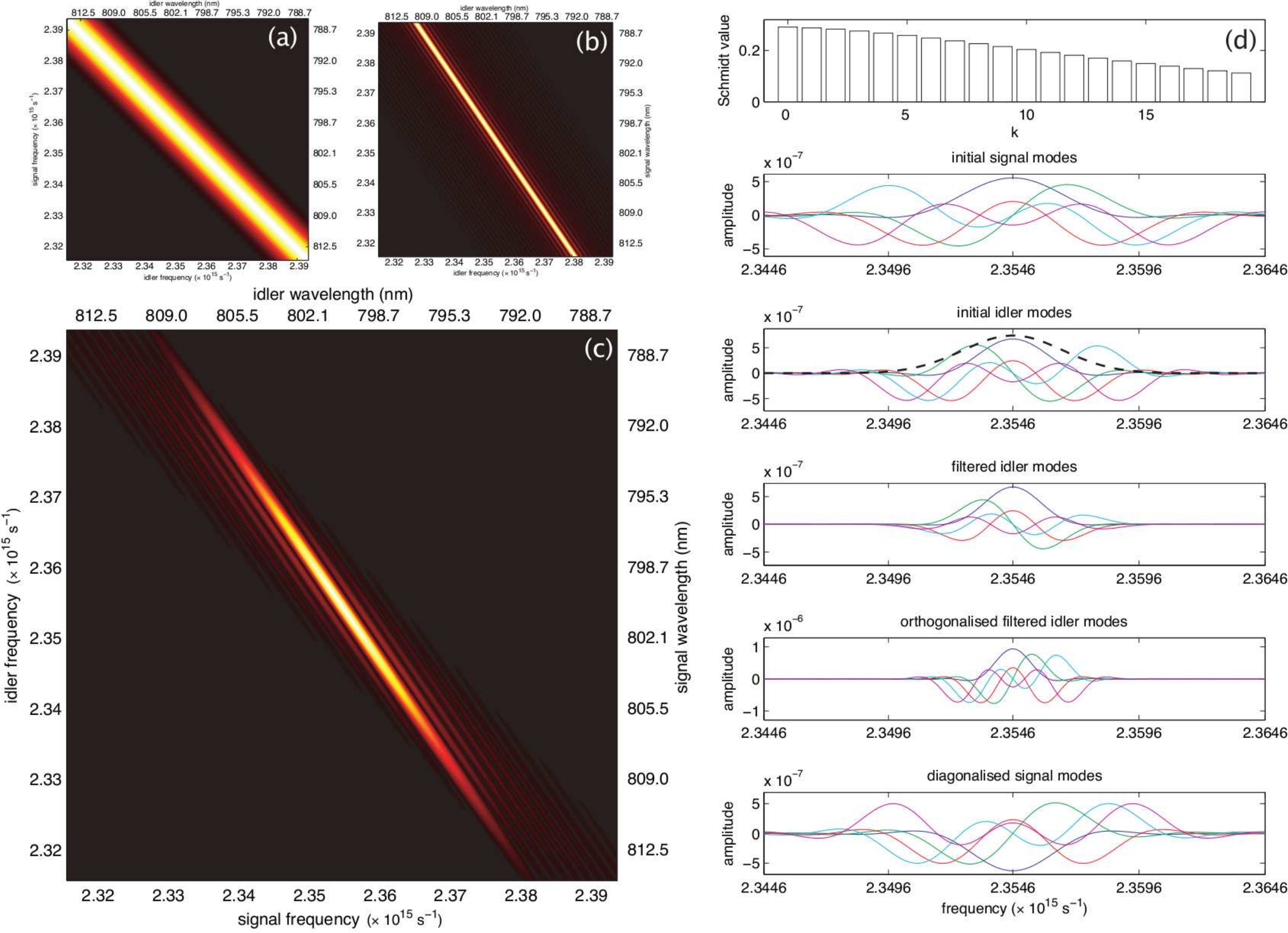} 
\end{center}
  \caption{(a) Gaussian pump function  $\alpha(\omega_i+\omega_s)$ with a 1nm FWHM at $\mu_p = 400$ nm.  (b) Phase matching function $\Phi(\omega_i,\omega_s)$ for waveguide of length $L=3.6$ mm and a periodicity of $\Lambda=8.8~\mu\mathrm{m}$.  (c) The resulting JSA $f(\omega_i,\omega_s)=\alpha(\omega_i+\omega_s)\Phi(\omega_i,\omega_s)$. \emph{The JSA has been plotted as a function of the frequency, however corresponding values for the wavelength have been included.} (d) Schmidt numbers and modes for the JSA (top to bottom): the first 20 Schmidt numbers $b_k$; the first 5 Schmidt modes $\xi_k(\omega_s)$ for the signal state; the first 5 Schmidt modes $\zeta_k(\omega_i)$ for the ilder state, as well as a Gaussian filter function of width $\sigma_f =2$$\times 10^{12}~\mathrm{s}^{-1}$ (dashed line); the filtered Schmidt modes $T(\omega_i)\zeta_k(\omega_i)$ for the idler state; the othogonalised idler modes $\phi_j(\omega_i)$; the diagonalised signal modes $\tau_m(\omega_s)$}
  \label{fig:JSA}
\end{figure}

As a physical example, we model a type II PP-KTP waveguide of length $L=3.6$ mm and a periodicity of $\Lambda=8.8~\mu\mathrm{m}$, pumped with a $400$ nm laser with a 1nm FWHM ($\sigma_p=5.00$$\times 10^{12}~\mathrm{s}^{-1}$) which down converts to $800$ nm in the signal and idler modes.  In figure \ref{fig:JSA} (a)-(c), we have plotted the pump function, the PMF and the JSA for the given parameters. Figure \ref{fig:JSA} (d) shows the corresponding Schmidt modes: \emph{initial signal modes} and \emph{initial idler modes}. In addition, it can be seen that after filtering the idler state, the idler Schmidt modes take on different spectral shapes. These \emph{filtered idler modes} are no longer orthogonal to each other and therefore need to be orthogonalised giving the \emph{orthogonalised idler modes}. When the idler mode is detected, the signal state gets projected into a mixture of orthogonal modes, as shown by the \emph{diagonalised signal modes}. Filtering and detection of the idler state changes the spectral shape of the signal state even though there is no physical interaction. This is a typical effect of entanglement. 

In general, the Schmidt decomposition can not be found analytically, but can be calculated numerically by computing the singular value decomposition of a discretised JSA. Unless stated otherwise, the results in this section were obtained using an $800\times 800$ grid, ranging over $0.2 \times 10^{15}$ $\mathrm{s}^{-1}$, centered around $\omega_i=\omega_s=\mu$. We note that an insufficiently fine grid, or insufficiently large region, will result in inflated values for the purity. The entropy of entanglement for this particular JSA is $E=4.6$. 

In this section we present results for: the probability of detecting a single photon in the idler mode; the $g^{(2)}$ and purity of the heralded state in the signal mode; and the fidelity between the signal state and the desired ideal Fock state.  We compare results for: an unfiltered idler state; an idler state filtered with a Gaussian filter $T(\omega_i)=\exp(-(\mu_f-\omega_i)/2\sigma_f^2)$, of various widths $\sigma_f $ and centered at the central idler frequency, where the filter function has been scaled such that the maximum value is always $1$; as well as the limiting case where $T(\omega_i)=\delta(\omega_i-\mu_f)$. We also present similar results for heralding a two-photon state conditional on the detection of two photons in the idler mode.

\subsubsection{Generating single-photon Fock states}
Due to the second-order truncation of the PDC output state, we are not considering 6- (or higher) photon contributions. At $\chi=0.5$, the fraction of 6-photon states, to 2-photon states, would roughly be $\chi^6/\chi^2=1/16$. We will not plot results beyond that. 

\begin{figure}[b!]
 \begin{center}
  \vspace{-.2cm}
   \includegraphics[width=7cm]{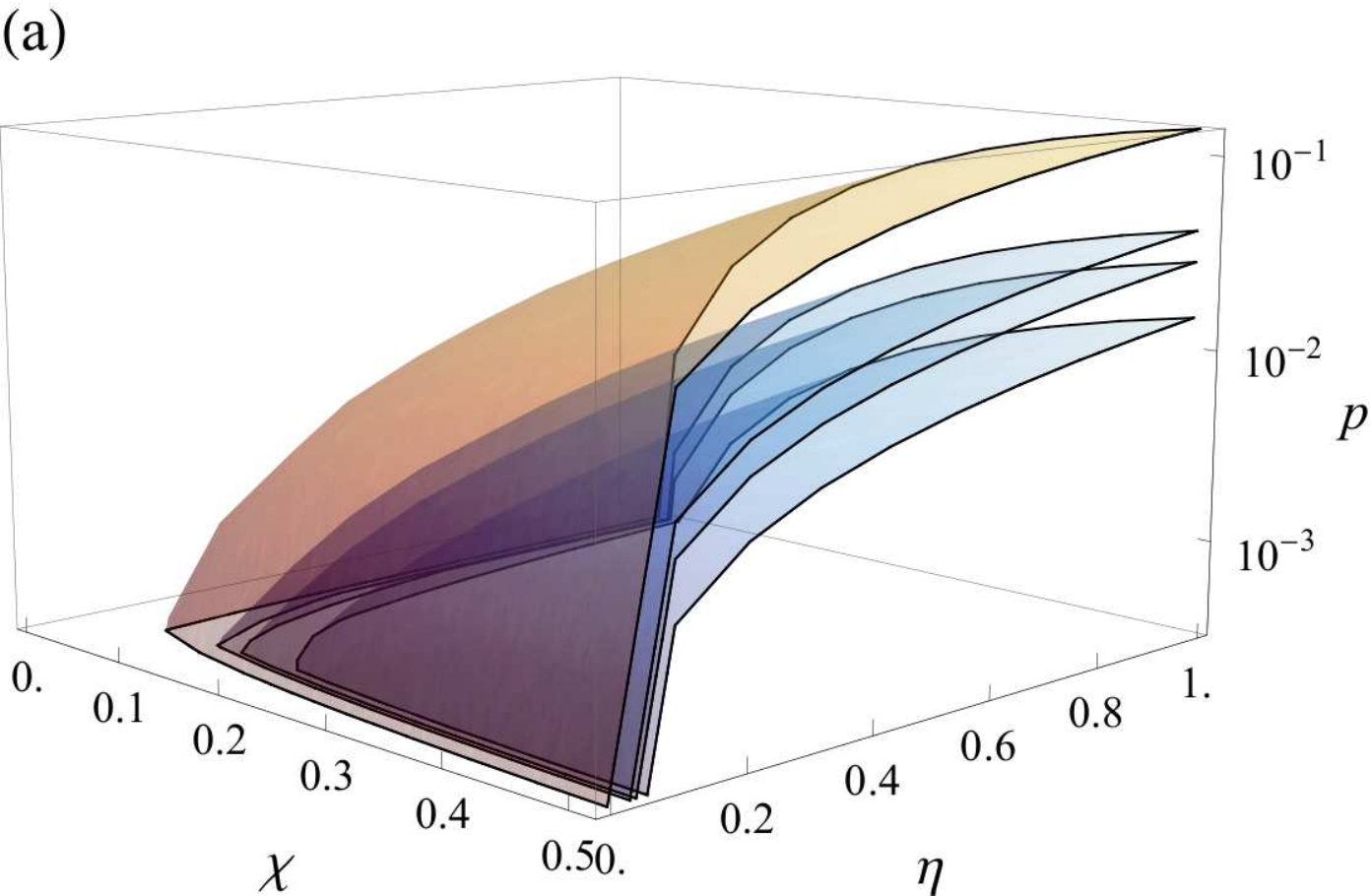} ~~~     \includegraphics[width=7cm]{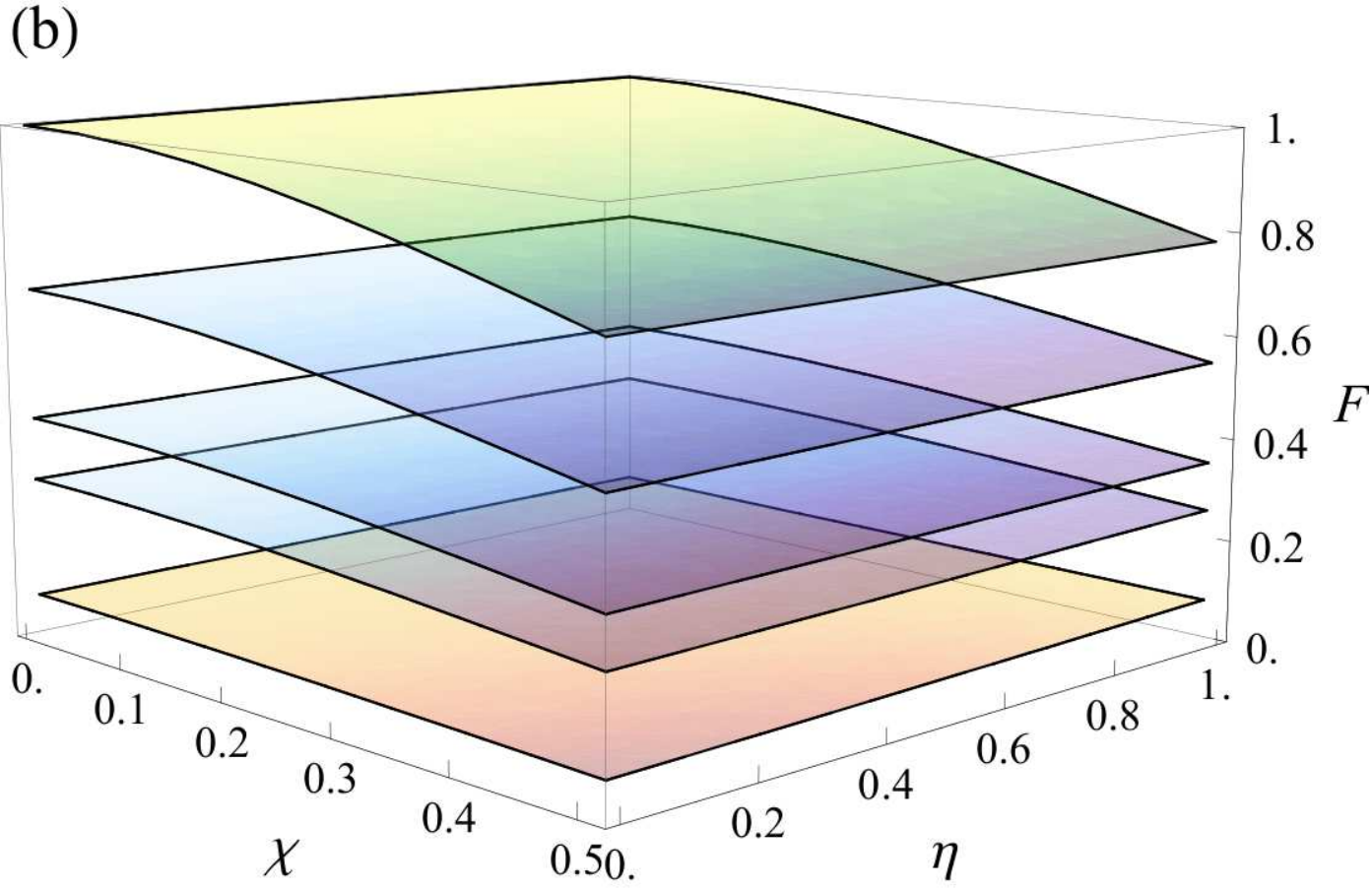}\\
         \includegraphics[width=7cm]{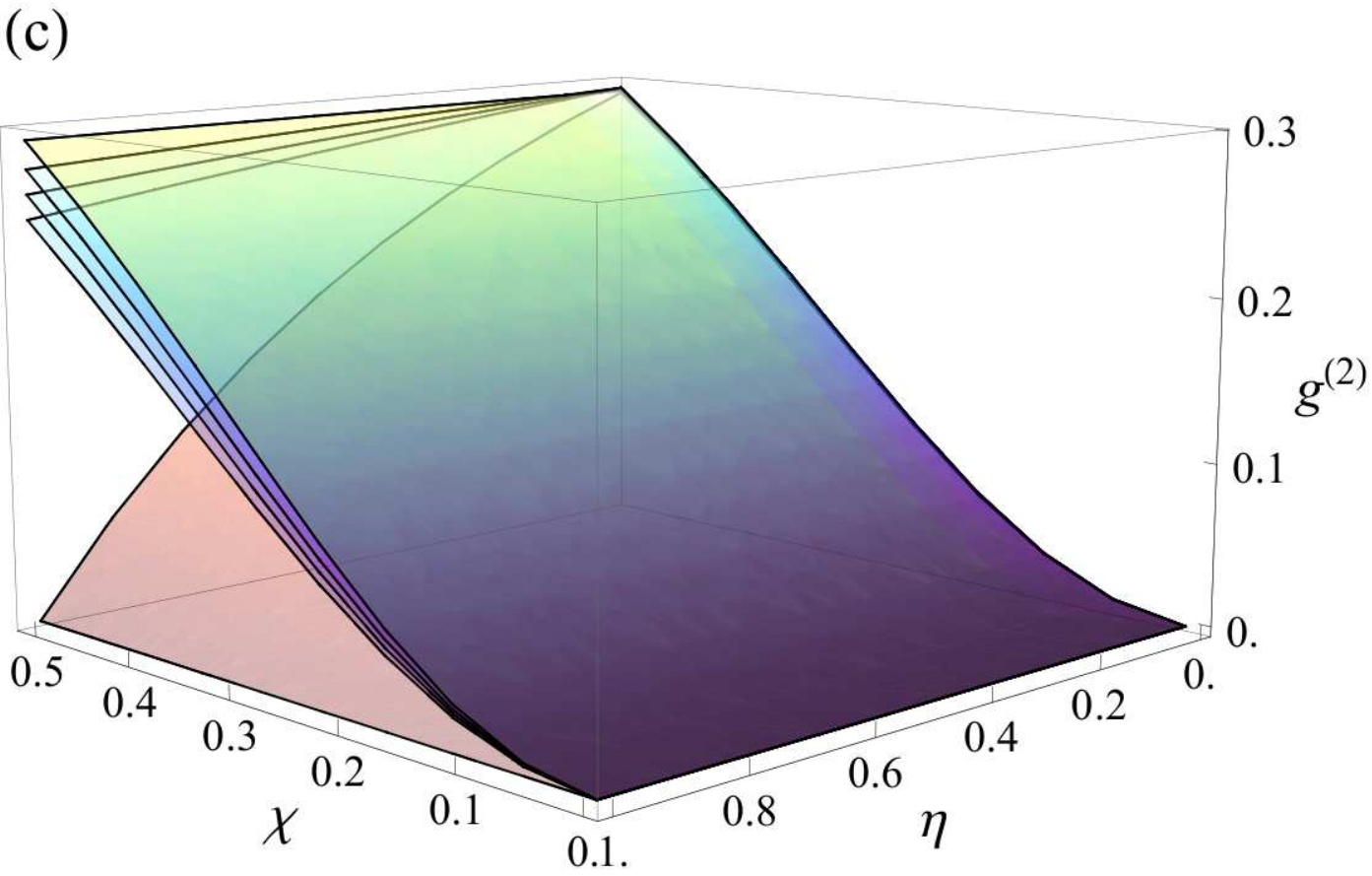}  ~~~ \includegraphics[width=7cm]{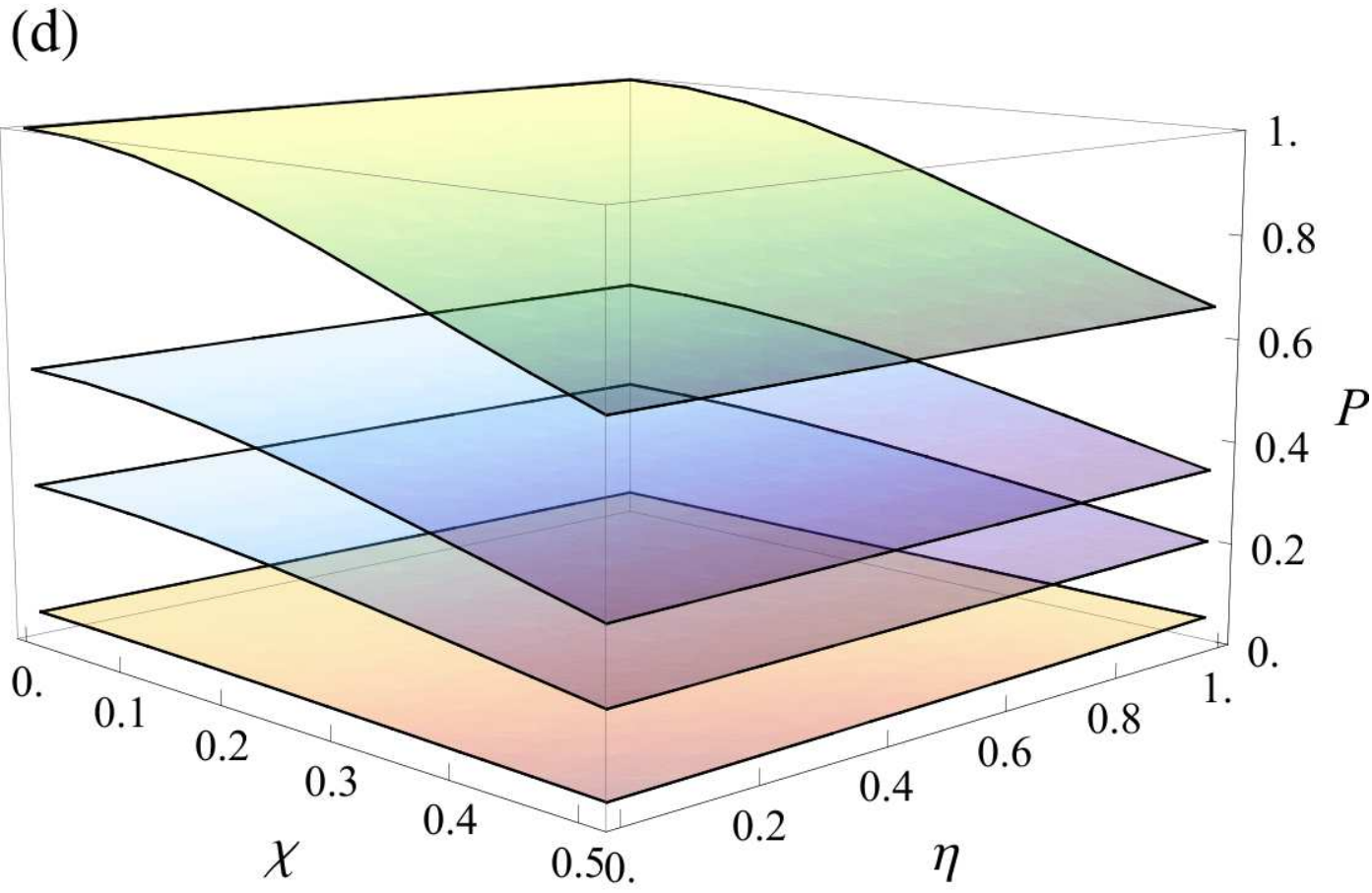}
   \vspace{-.5cm}
\end{center}
  \caption{(a) The probability of detecting a single photon in the idler mode for (top to bottom): no filter; $\sigma_f =3$$\times 10^{12}~\mathrm{s}^{-1}$; $\sigma_f =2$$\times 10^{12}~\mathrm{s}^{-1}$; $\sigma_f =1$$\times 10^{12}~\mathrm{s}^{-1}$. \emph{Note that the probability is plotted on a log scale. } (b)  The fidelity of the signal state with an ideal Fock state for (top to bottom): $\sigma_f =0$; $\sigma_f =1$$\times 10^{12}~\mathrm{s}^{-1}$; $\sigma_f =2$$\times 10^{12}~\mathrm{s}^{-1}$; $\sigma_f =3$$\times 10^{12}~\mathrm{s}^{-1}$; no filter. (c) The $g^{(2)}$ of the signal state for (top to bottom): $\sigma_f =0$; $\sigma_f =1$$\times 10^{12}~\mathrm{s}^{-1}$; $\sigma_f =2$$\times 10^{12}~\mathrm{s}^{-1}$; $\sigma_f =3$$\times 10^{12}~\mathrm{s}^{-1}$; no filter. \emph{Note the change in axis orientation.} (d) The purity of the signal state for (top to bottom): $\sigma_f =0$; $\sigma_f =1$$\times 10^{12}~\mathrm{s}^{-1}$; $\sigma_f =2$$\times 10^{12}~\mathrm{s}^{-1}$; no filter.  } 
  \label{fig:probfid1} 
\end{figure}

In figure \ref{fig:probfid1} (a), we have plotted the probability of detecting a single photon in the idler mode as a function of the nonlinearity $\chi$ and the efficiency of the detector $\eta$.  Notice that the probability of detecting a single photon in the idler mode increases with higher detector efficiency and higher nonlinearity strength, as expected. 

The fidelity has been plotted in figure \ref{fig:probfid1} (b). The inclusion of a filter has a drastic effect on the fidelity. It has a greater dependence on the strength of the nonlinearity, than in the unfiltered case, however the overall fidelity is much higher. Notice as well that there is a trade off between the fidelity and the probability of detection. 

In figure \ref{fig:probfid1} (c), we have plotted the $g^{(2)}$ for the heralded state in the signal mode. Note that, for visual clarity, the figure orientation has been rotated by $\pi$ around the $z{-}\mathrm{axis}$, with respect to the other plots. It is useful to know that $g^{(2)}=0$ for a single-photon state and $1/2$ for a two-photon state. A curious thing is that decreasing the filter width results in higher fidelities, despite the higher proportion of two-photon states, as shown by the $g^{(2)}$. This suggests that for this particular JSA, the dominant cause of impurity is the spectral entanglement, rather than the resulting photon-number mixture due to the presence of higher-order terms.

In figure \ref{fig:probfid1} (d), we have plotted the purity of the state in the signal mode. Due to the 4-fold summation, the purity for the filtered case is very demanding computationally, therefore, we have only included examples of two filter widths. Results for the filtered case were computed using a $600\times 600$ grid, ranging over $0.16 \times 10^{15}$ $\mathrm{s}^{-1}$, centered around $\omega_i=\omega_s=\mu$, and truncating $b_k$ with values below $10^{-2}$.

\begin{figure}[h!]
 \begin{center}
  \vspace{-.2cm}
   \includegraphics[width=5.1cm]{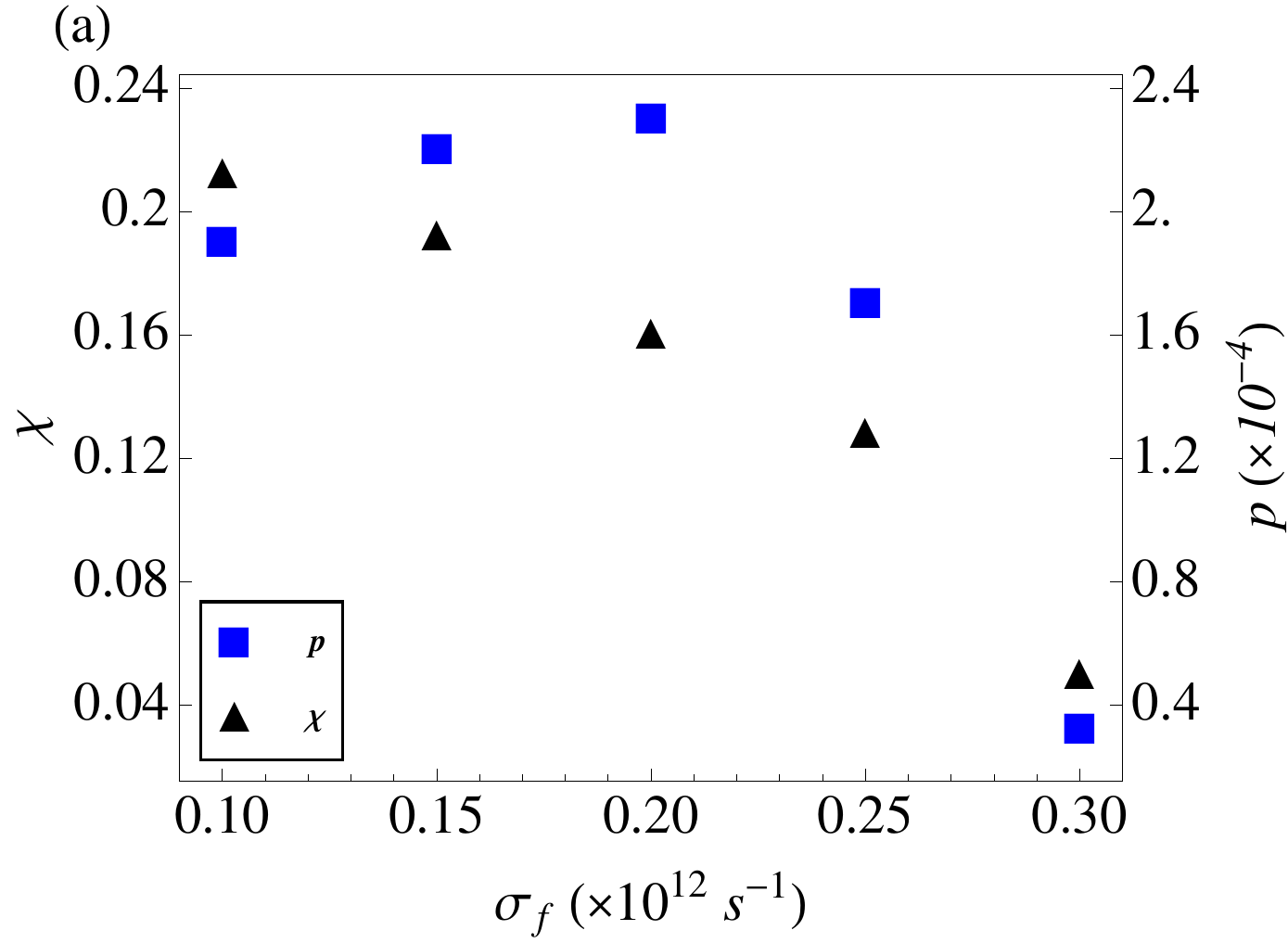}   \includegraphics[width=5.1cm]{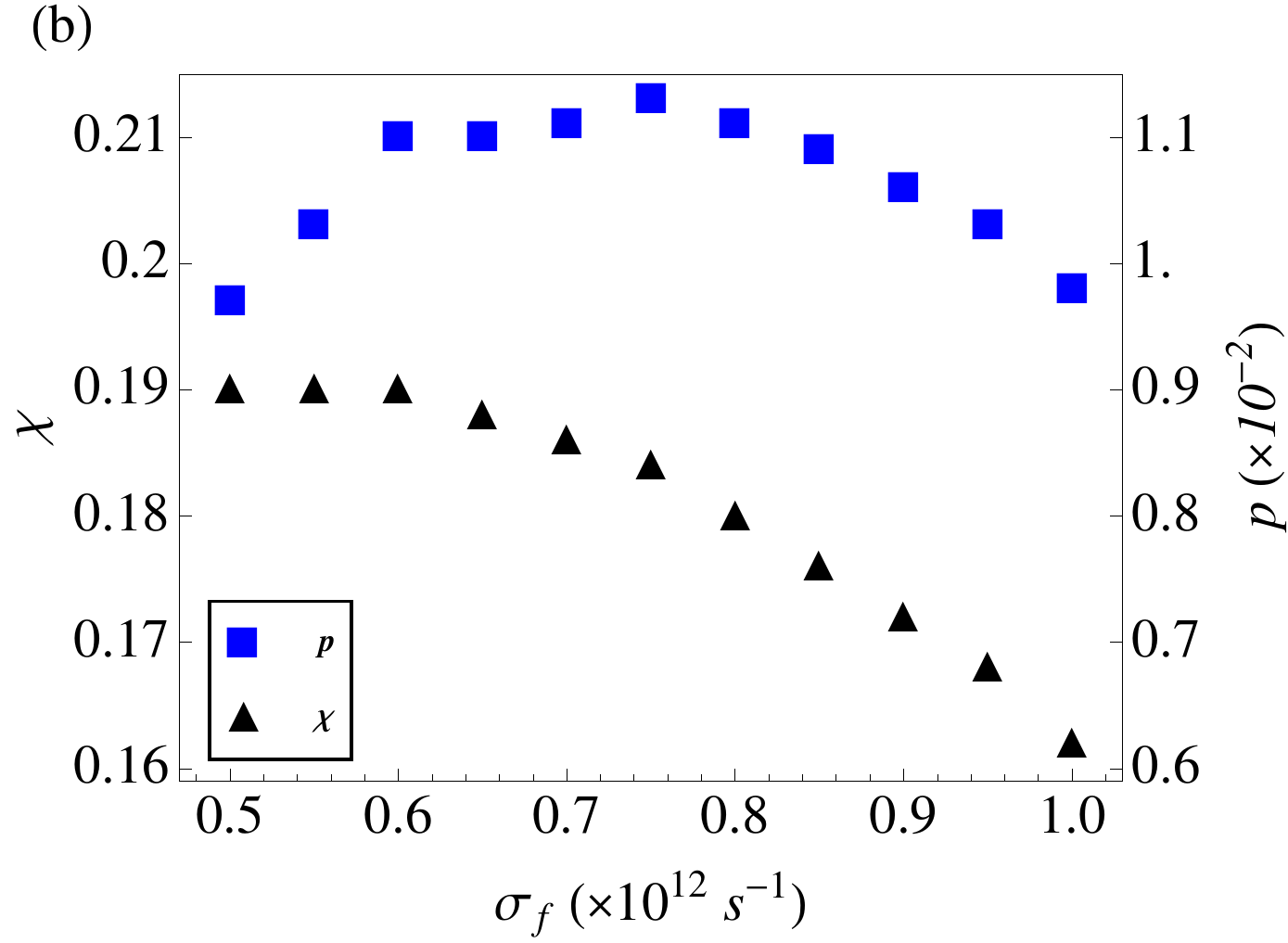}
         \includegraphics[width=5.1cm]{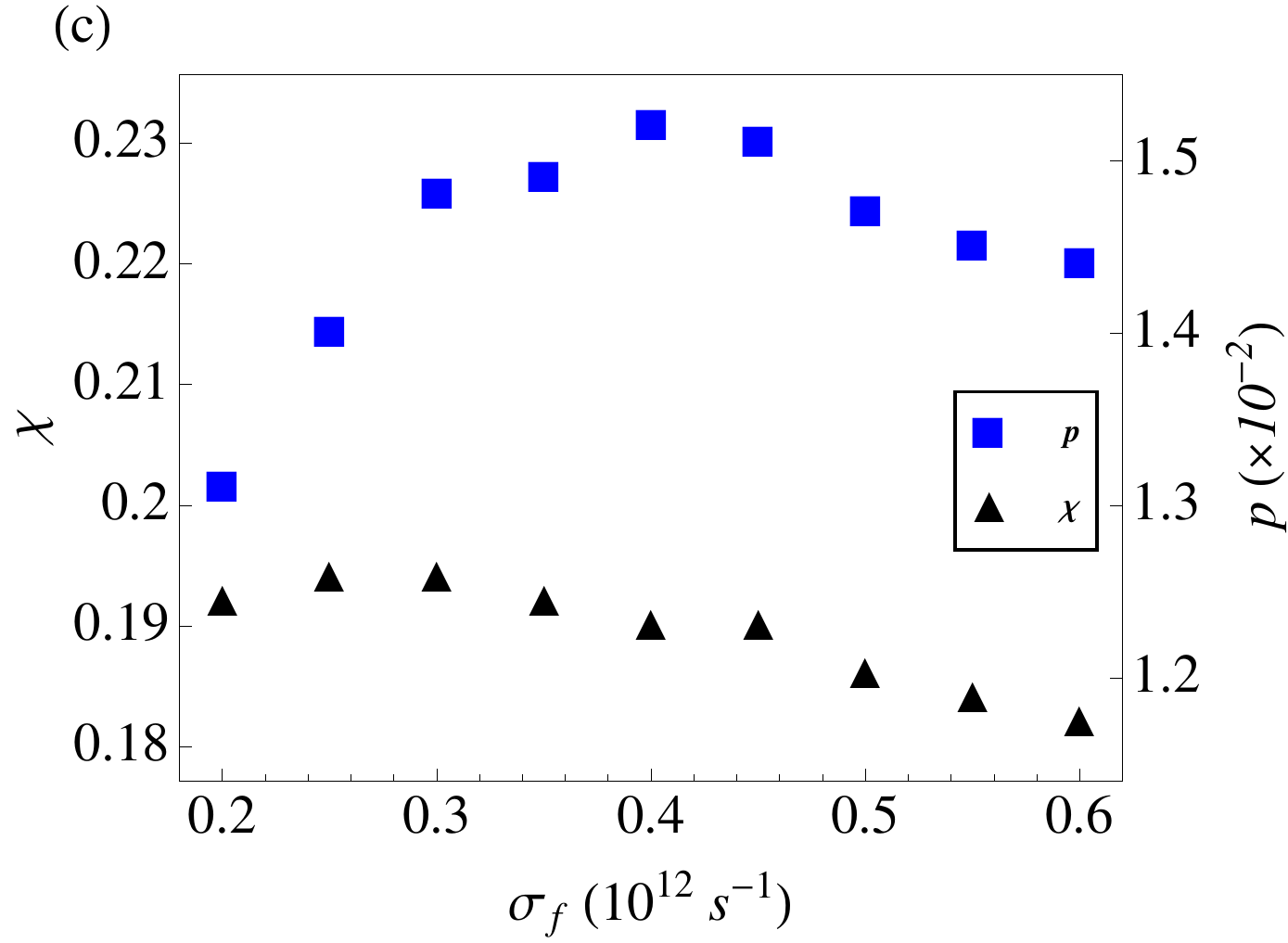}  
   \vspace{-.5cm}
\end{center}
  \caption{ Probability of detecting a single photon in the idler mode, and required nonlinearity, for a number of filter widths, in order to achieve a fidelity of $F=0.95$, using a heralding detector with efficiency $\eta=0.5$, for: (a) correlated JSA  (see section \ref{sec:phys1}); (b) symmetric JSA (see section \ref{sec:sym}); (c) asymmetric JSA  (see section \ref{sec:phys3}).} 
  \label{fig:ChiProb} 
\end{figure}

To achieve a fidelity of $F=0.95$, using a heralding detector with efficiency $\eta=0.5$, we could choose from a range of filter widths at different nonlinearity strengths. Different combinations, however, result in slightly different probabilities of success. Figure \ref{fig:ChiProb}~(a) shows the probability of success, and required nonlinearity, for a number of filter widths.

\subsubsection{Generating 2-photon Fock states}

\begin{figure}[h!]
  \begin{center}
 \includegraphics[width=7cm]{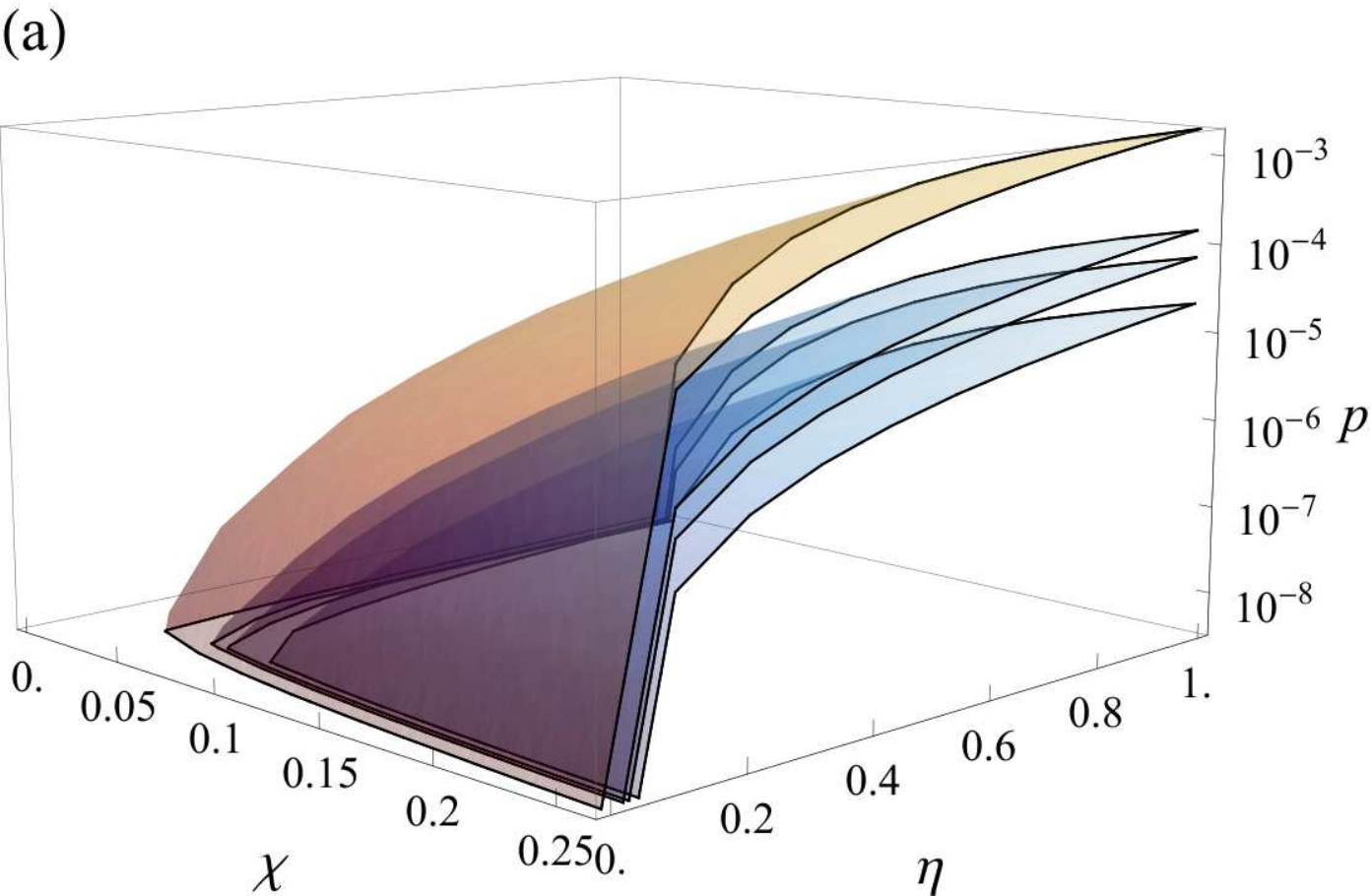} ~~~     \includegraphics[width=7cm]{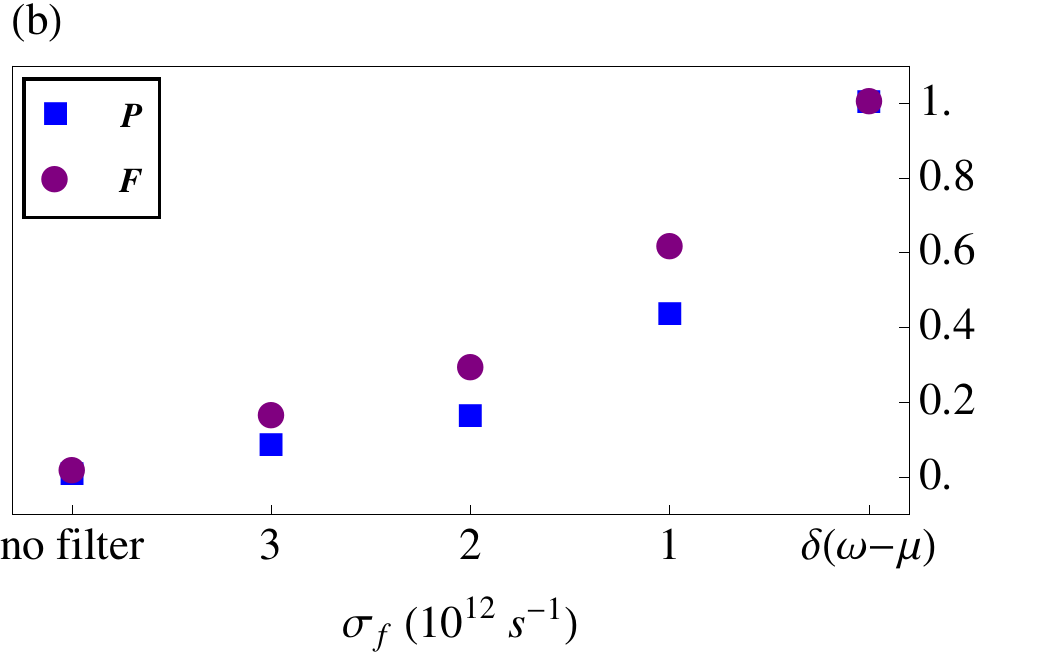}
   \vspace{-.5cm}
  \end{center}
  \caption{(a) The probability of detecting two photons in the idler mode for (top to bottom): no filter; $\sigma_f =3$$\times 10^{12}~\mathrm{s}^{-1}$; $\sigma_f =2$$\times 10^{12}~\mathrm{s}^{-1}$; $\sigma_f =1$$\times 10^{12}~\mathrm{s}^{-1}$. \emph{Note that the probability is plotted on a log scale. }(b)  The purity and fidelity of the signal state with an ideal Fock state. } 
  \label{fig:prob2}
\end{figure}

At $\chi=0.25$, the fraction of 6-photon states, to 4-photon states, would roughly be $\chi^6/\chi^4=1/16$. We will not plot results beyond that. Figure \ref{fig:prob2} (a) shows the probability for an inefficient detector to detect two photons in the idler mode. The corresponding fidelities and purities have been shown in figure \ref{fig:prob2} (b) and do not vary as a function of $\chi$ and $\eta$.

\section{Physical example II - group velocity matching}\label{sec:phys2}

In this section, we will examine particular phase matching conditions which result in a less entangled JSA, and therefore a more pure heralded Fock state. It is common to approximate the phasematching function as $\Phi(\omega_i,\omega_s)=\exp(-\gamma L^2\Delta k^2/4)$ where $\gamma\approx 0.193$. By making this approximation, we can write the JSA as
\begin{eqnarray}\label{eq:JSA_sep}
f(\omega_i,\omega_s)\propto\exp\Big({-}\frac{(\omega_i+\omega_s-2\mu)^2}{2\sigma_p^2}\Big)\exp\Big({-} \frac{ \gamma L^2\Delta k^2}{4}\Big)\,.
\end{eqnarray}
In order to make equation (\ref{eq:JSA_sep}) separable, we require all ``cross-terms'', i.e. terms which contain products of $\omega_i$ and $\omega_s$ to vanish. This occurs when the condition
\begin{eqnarray}\label{eq:special_cond}
\frac{2}{\sigma^2}+\gamma L^2(k_s'-k_p')(k_i'-k_p')=0
\end{eqnarray}
is met, yielding a JSA of the form $f(\omega_i,\omega_s)\propto f_i(\omega_i)f_s(\omega_s)$ \cite{URen2005}. One way to satisfy the condition in equation (\ref{eq:special_cond}) is to set $k_p'=(k_s'+k_i')/2$, which results in the following condition for the length of the waveguide, as a function of the pump width:
\begin{eqnarray}
L=1/\sqrt{8\gamma \sigma_p^2(k_s'-k_i')^2}\,.
\end{eqnarray}
These conditions generate a symmetric JSA, where both signal and idler modes have equal widths. Alternatively, rearranging equation (\ref{eq:special_cond}) as follows
\begin{eqnarray}\label{eq:special_cond2}
\frac{4}{\sigma L(k_i'-k_p')}+\gamma \sigma L(k_s'-k_p')=0\,,
\end{eqnarray}
we can see that by making $L<<\sigma^{-1}$, i.e. $L\rightarrow \infty$ and setting $k_p'=k_s'$, we can also obtain a separable JSA. These conditions generate an asymmetric JSA \cite{URen2005}.

We have made use of the Gaussian approximation for $\Phi(\omega_i,\omega_s)$ to obtain the conditions for separability, however we will now input these conditions into the original sinc form of the function. This analysis will not result in completely pure states being generated, however it should correspond more closely to experimental observations. 

\subsection{Symmetric JSA}\label{sec:sym}

In order to meet the extended phase matching conditions for a symmetric, separable JSA, we again model a type II PP-KTP waveguide, now of length $L=24.2$ mm and a periodicity of $\Lambda=68.4~\mu\mathrm{m}$, pumped with a $788$ nm laser with a 0.7nm FWHM ($\sigma_p=0.9$$\times 10^{12}~\mathrm{s}^{-1}$) which down converts to $1576$ nm in the signal and idler modes. Unless stated otherwise, the results in this section were obtained using an $800\times 800$ grid, ranging over $0.06 \times 10^{15}$ $\mathrm{s}^{-1}$, centered around $\omega_i=\omega_s=\mu$.

\begin{figure}[t!]
\includegraphics[width=16cm]{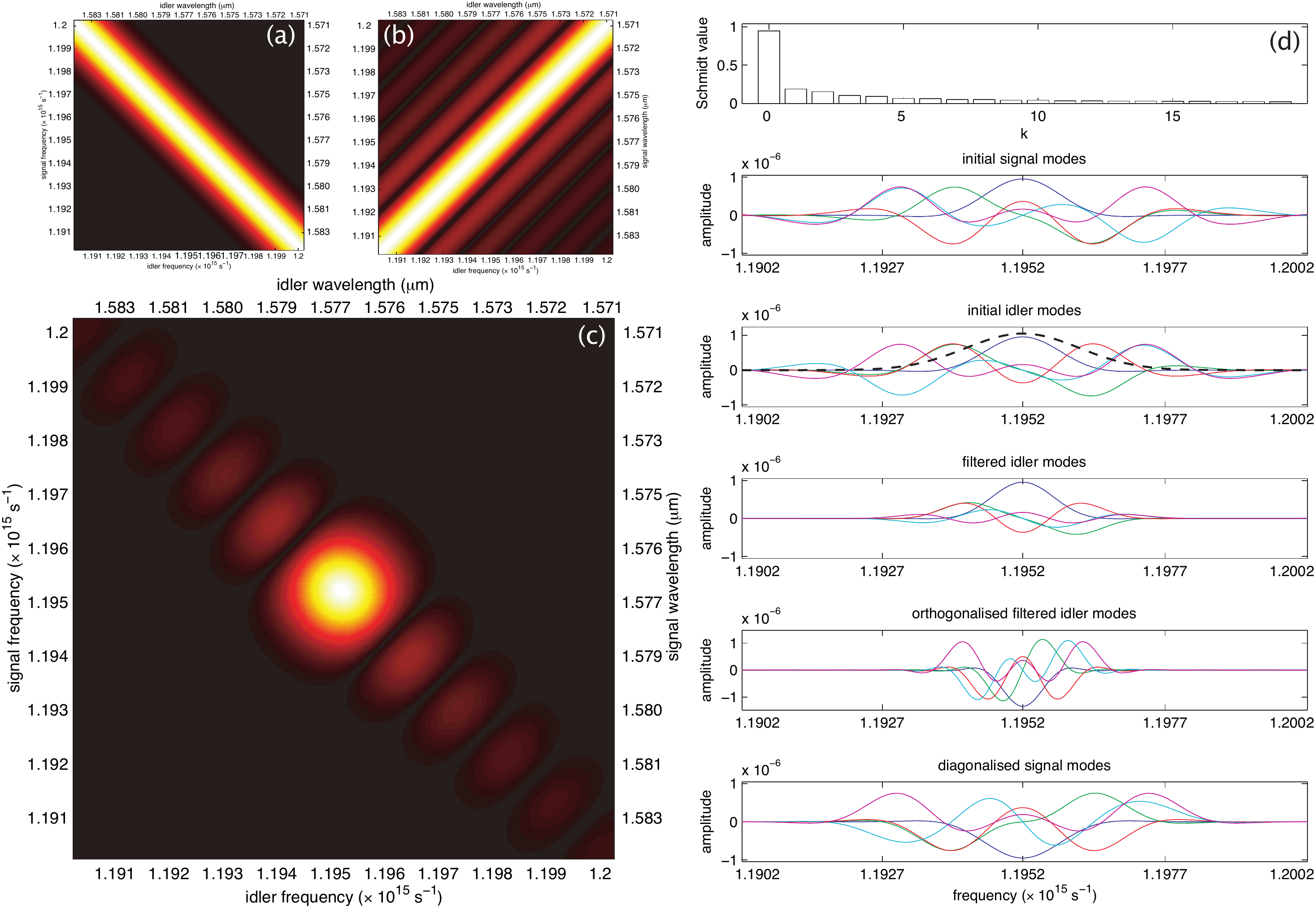}
    \caption{  (a) Gaussian pump function  $\alpha(\omega_i+\omega_s)$ with a 0.7nm FWHM at $\mu_p = 788$ nm. (b) Phase matching function $\Phi(\omega_i,\omega_s)$ for waveguide of length $L=24.2$ mm and a periodicity of $\Lambda=68.4~\mu\mathrm{m}$. (c) The resulting JSA $f(\omega_i,\omega_s)=\alpha(\omega_i+\omega_s)\Phi(\omega_i,\omega_s)$. \emph{The JSA has been plotted as a function of the frequency, however corresponding values for the wavelength have been included.} (d) Schmidt numbers and modes for the JSA (top to bottom): the first 20 Schmidt numbers $b_k$; the first 5 Schmidt modes $\xi_k(\omega_s)$ for the signal state; the first 5 Schmidt modes $\zeta_k(\omega_i)$ for the ilder state, as well as a Gaussian filter function of width $\sigma_f =1$$\times 10^{12}~\mathrm{s}^{-1}$ (dashed line); the filtered Schmidt modes $T(\omega_i)\zeta_k(\omega_i)$ for the idler state; the othogonalised idler modes $\phi_j(\omega_i)$; the diagonalised signal modes $\tau_m(\omega_s)$}
  \label{fig:JSA_ext}
\end{figure}

Figure \ref{fig:JSA_ext} shows the JSA and the corresponding Schmidt values and modes. Notice in the Schmidt decomposition that the first mode is much more dominant than it was in section \ref{sec:phys1}. The entropy of entanglement for this JSA is $E=0.88$. If we had used the Gaussian approximation for $\Phi$, the JSA would decompose into one pair of Schmidt modes and the entropy of entanglement would be $E=0$. In such a case, the four-photon term would would consist only of two-photon Fock states. 

As an intuitive guide to why the above conditions generate the given JSA, notice that varying the parameter $\Delta k$ has the effect of changing the gradient of the phase matching function $\Phi(\omega_i,\omega_s)$ (see figure  \ref{fig:JSA_ext} (b)), rotating it around $\omega_i=\omega_s=\mu$ while changing the parameter $L$, alters the width of the phase matching function. The goal is to pick $\Delta k$, and therefore $k_p'$, and $L$ such that the phase matching function is perpendicular, and of equal width, to the pump function.

As in the previous section, we present results for the probability, purity, $g^{(2)}$ and fidelity, for the heralding of one and two photon Fock states. For a realistic JSA, manipulating the phase-matching conditions can result in high purity of the heralded state, however, it doesn't reach unity. From figure \ref{fig:JSA_ext} (c), it can be seen that the outer lobes contribute to the spectral correlations and perhaps it is possible to increase the purity of the heralded state by filtering them out. Therefore, we will again compare results for: an unfiltered idler state; an idler state filtered with a Gaussian filter $T(\omega_i)=\exp(-(\mu_f-\omega_i)/2\sigma_f^2)$, of various widths $\sigma_f$ and centered at the central idler frequency; as well as the limiting case where $T(\omega_i)=\delta(\omega_i-\mu_f)$.

\subsubsection{Generating single-photon Fock states}

 \begin{figure}[t!]
 \begin{center}
  \vspace{-.2cm}
 \includegraphics[width=7cm]{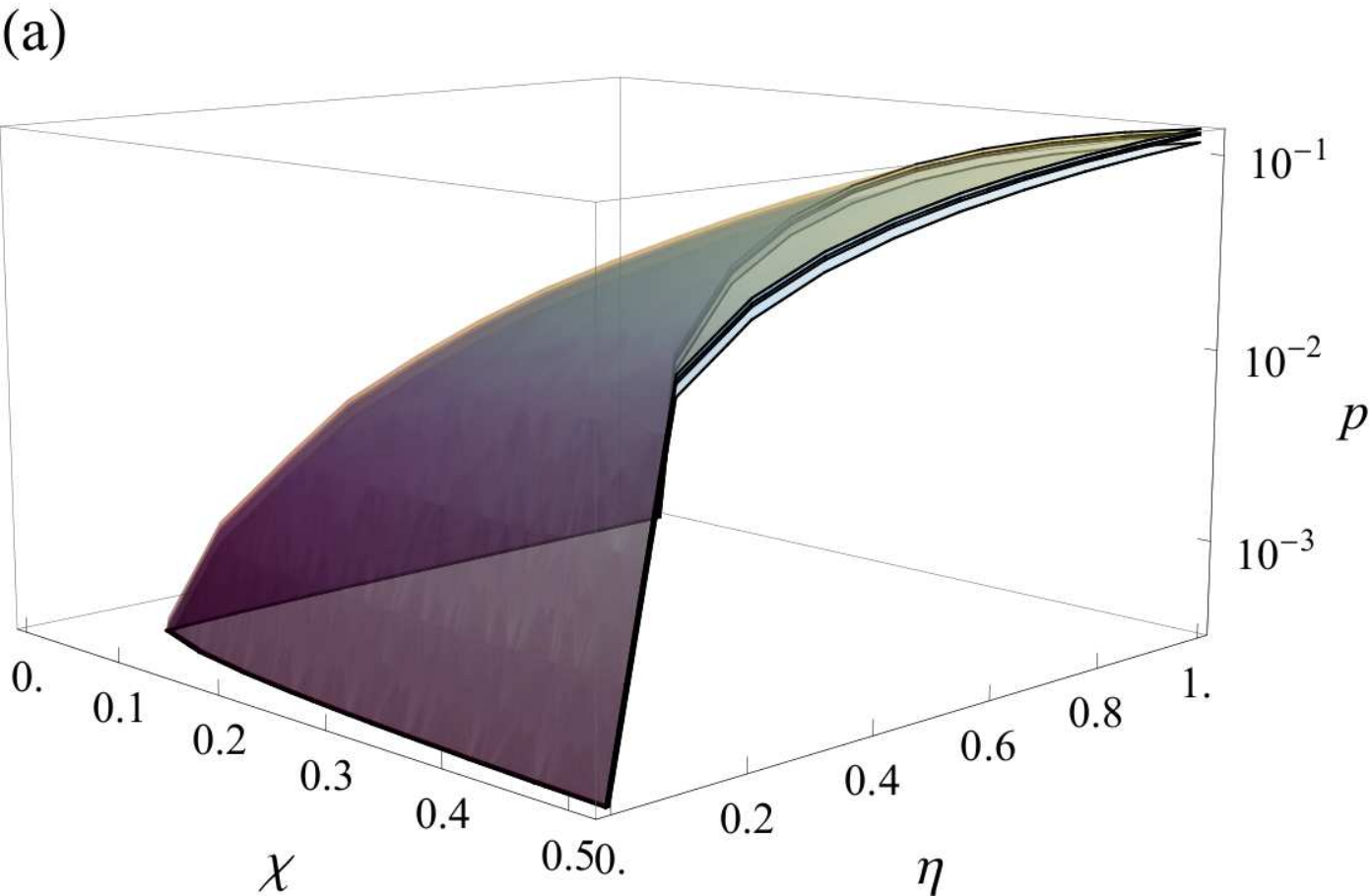} ~~~     \includegraphics[width=7cm]{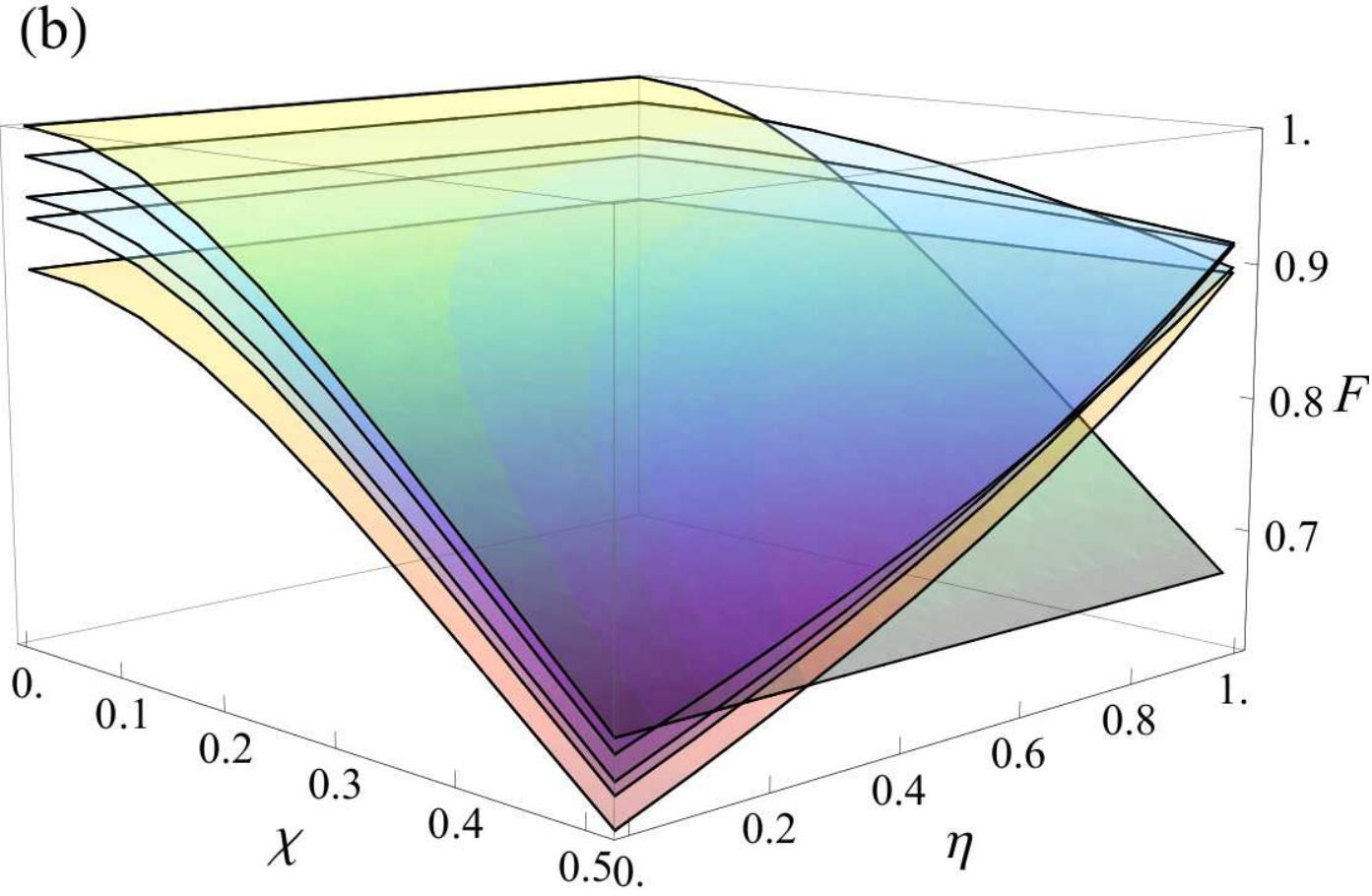}\\
         \includegraphics[width=7cm]{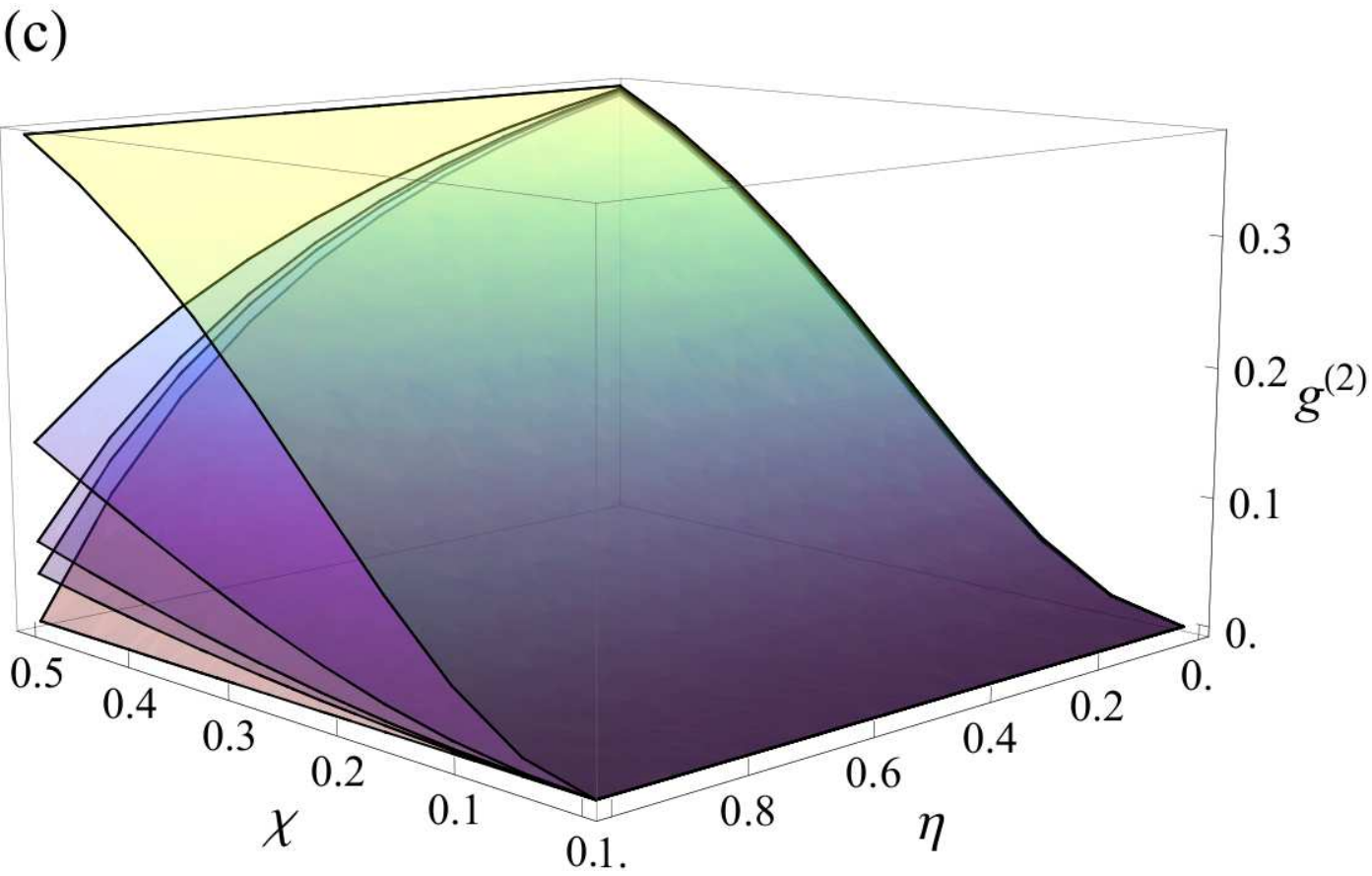}  ~~~ \includegraphics[width=7cm]{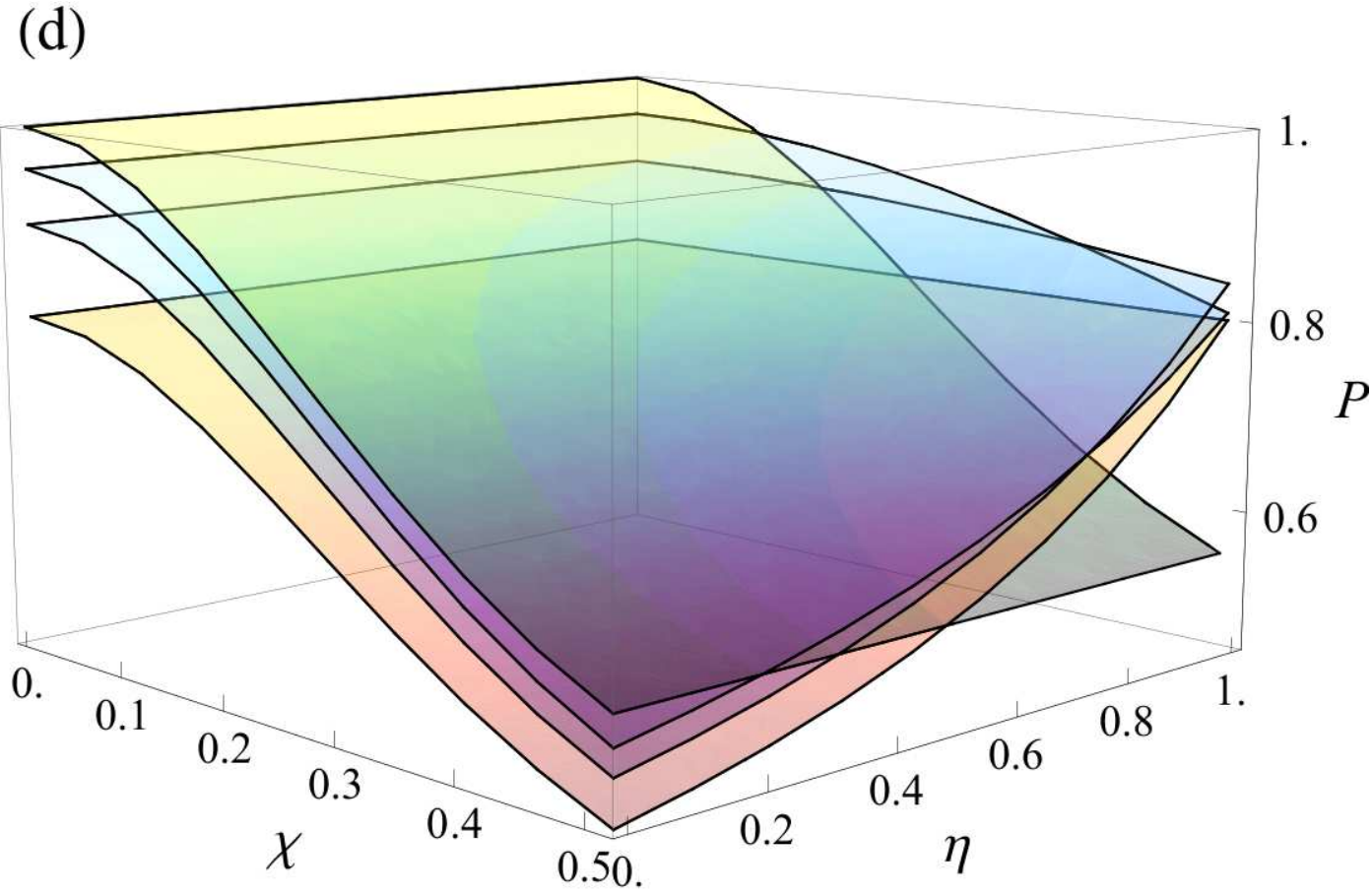}
   \vspace{-.5cm}
\end{center}
  \caption{(a) The probability of detecting a single photon in the idler mode for (top to bottom): no filter; $\sigma_f =3$$\times 10^{12}~\mathrm{s}^{-1}$; $\sigma_f =2$$\times 10^{12}~\mathrm{s}^{-1}$; $\sigma_f =1$$\times 10^{12}~\mathrm{s}^{-1}$. \emph{Note that the probability is plotted on a log scale. }(b)  The fidelity of the signal state with an ideal Fock state for (top to bottom at $\chi=0$ and $\eta=0$): $\sigma_f =0$; $\sigma_f =1$$\times 10^{12}~\mathrm{s}^{-1}$; $\sigma_f =2$$\times 10^{12}~\mathrm{s}^{-1}$; $\sigma_f =3$$\times 10^{12}~\mathrm{s}^{-1}$; no filter. (c) The $g^{(2)}$ of the signal state for (top to bottom): $\sigma_f =0$; $\sigma_f =1$$\times 10^{12}~\mathrm{s}^{-1}$; $\sigma_f =2$$\times 10^{12}~\mathrm{s}^{-1}$; $\sigma_f =3$$\times 10^{12}~\mathrm{s}^{-1}$; no filter. \emph{Note the change in axis orientation.} (d) The purity of the signal state for (top to bottom at $\chi=0$ and $\eta=0$): $\sigma_f =0$; $\sigma_f =1$$\times 10^{12}~\mathrm{s}^{-1}$; $\sigma_f =2$$\times 10^{12}~\mathrm{s}^{-1}$;  no filter. } 
  \label{fig:probfid1_ext} 
\end{figure}

Figure \ref{fig:probfid1_ext} (a) shows the probability of detecting a single photon in the signal mode. Since most of the photons will have spectral distributions within the filter width, we do not see a very big drop in the probability, when filtering. 

Figure \ref{fig:probfid1_ext} (b) shows the fidelity. We distinguish between these surfaces by referring to their values at the point $\chi=0$ and $\eta=0$.  Immediately we can see that the fidelity is much higher than in section~\ref{sec:phys1}. In the region of interest, filtering the idler mode increases the fidelity of the signal state with a single photon. Decreasing the filter width can be detrimental to the fidelity when the nonlinearity and the detector efficiency are high.

Figure \ref{fig:probfid1_ext} (c) shows the $g^{(2)}$ and figure~\ref{fig:probfid1_ext} (d) shows the purity of the state in the signal mode.  Again, we will distinguish between these surfaces by referring to their values at the point $\chi=0$ and $\eta=0$. Results for the filtered case were computed using a $600\times 600$ grid, ranging over $0.06 \times 10^{15}$ $\mathrm{s}^{-1}$, centered around $\omega_i=\omega_s=\mu$, and truncating $b_k$ with values below $10^{-2}$. 

Figure \ref{fig:ChiProb}~(b) shows the probability of success, and required nonlinearity, for a number of filter widths, in order to achieve a fidelity of $F=0.95$, using a detector with efficiency $\eta=0.5$. Note that while the probability of success is greatly enhanced by using a source engineered state, the required pump powers are nearly the same. Also notice the ``flat'' region, where the probability does not change much, between $\sigma_f=0.6\times10^{12}~\mathrm{s}^{-1}$ and  $0.7\times10^{12}~\mathrm{s}^{-1}$. This corresponds to the ``dark'' region between the lobes on the JSA. Over this region, we do not expect much change in the flux.

\subsubsection{Generating 2-photon Fock states}

\begin{figure}[h!]
  \begin{center}
 \includegraphics[width=7cm]{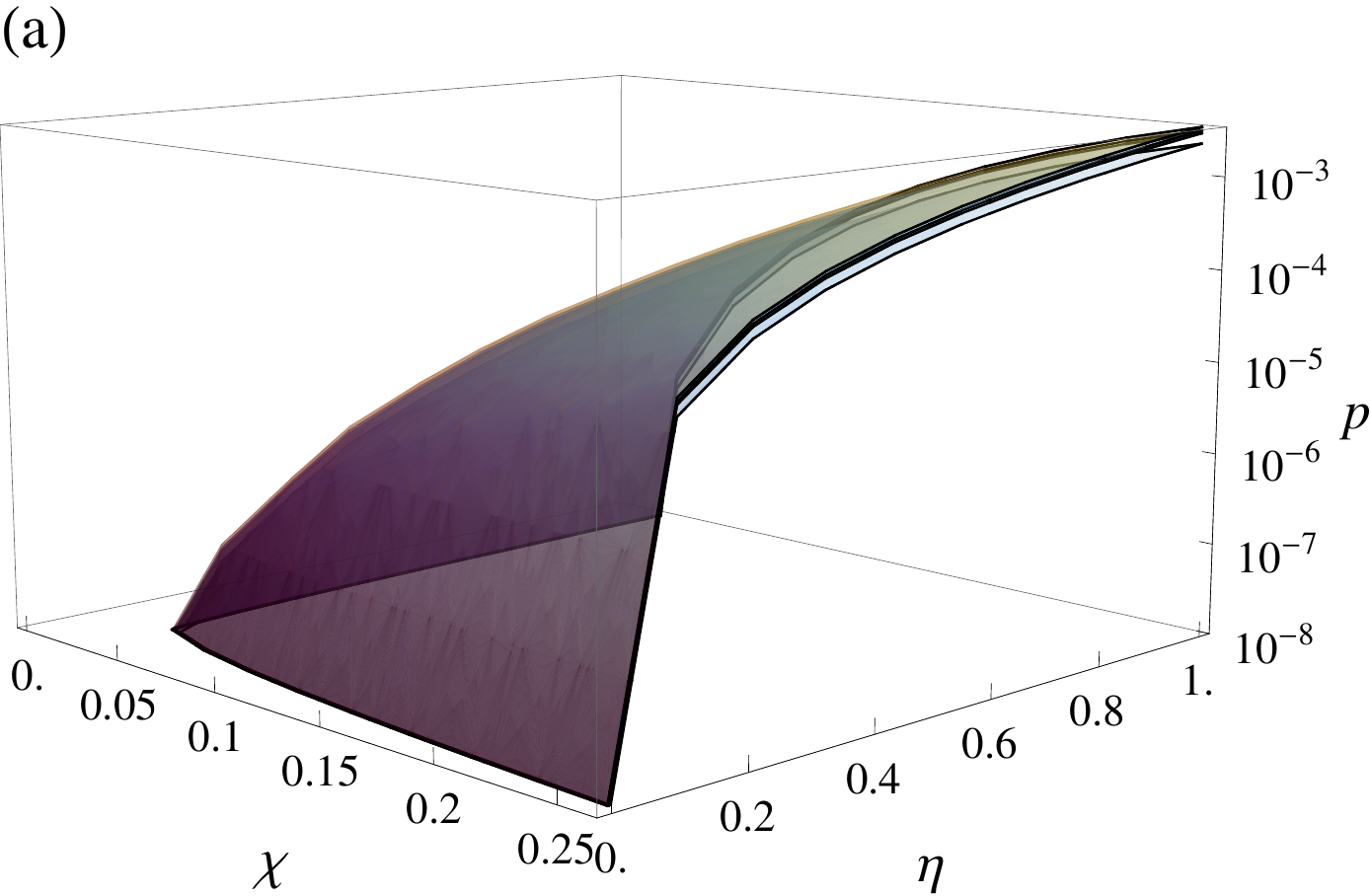} ~~~     \includegraphics[width=7cm]{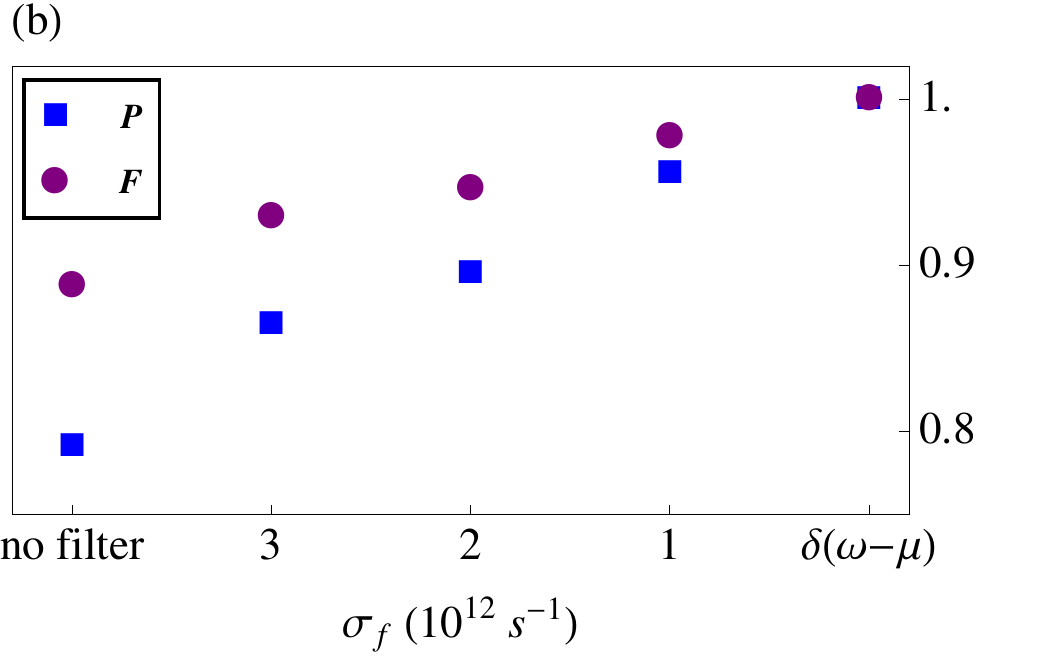}
   \vspace{-.5cm}
  \end{center}
    \caption{(a) The probability of detecting two photons in the idler mode for (top to bottom): no filter; $\sigma_f =3$$\times 10^{12}~\mathrm{s}^{-1}$; $\sigma_f =2$$\times 10^{12}~\mathrm{s}^{-1}$; $\sigma_f =1$$\times 10^{12}~\mathrm{s}^{-1}$. \emph{Note that the probability is plotted on a log scale. }(b)  The purity and fidelity of the signal state with an ideal Fock state. } 
  \label{fig:prob2ext}
\end{figure}

Figure \ref{fig:prob2ext} (a), shows the probability for an inefficient detector to detect two photons in the idler mode. The corresponding fidelities and purities have been shown in figure \ref{fig:prob2ext} (b). The fidelity and purity do not vary as a function of $\chi$ and $\eta$.

\subsection{Asymmetric JSA}\label{sec:phys3}

\begin{figure}[t!]
\includegraphics[width=16cm]{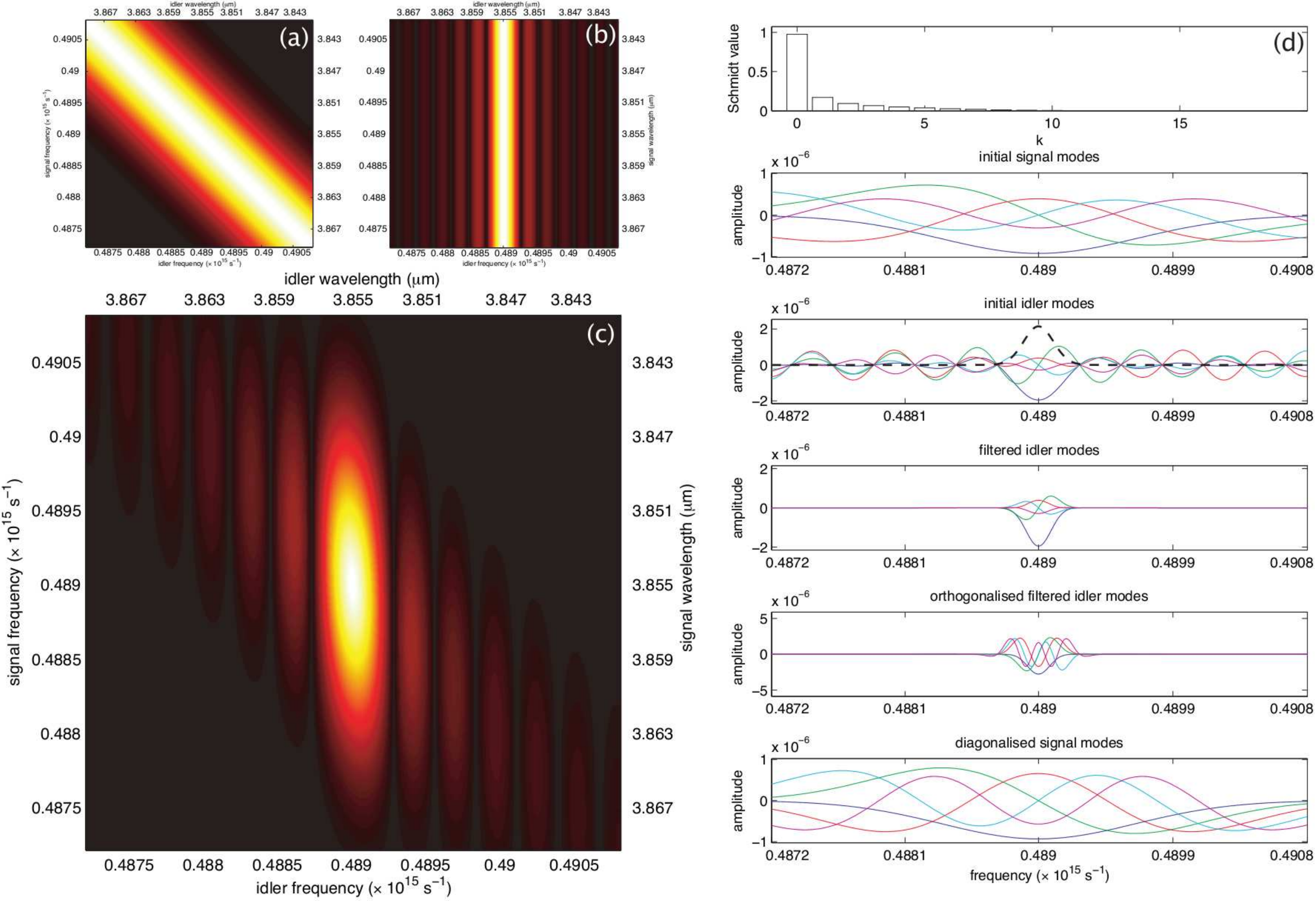}
      \caption{ (a) Gaussian pump function  $\alpha(\omega_i+\omega_s)$ with a 3nm FWHM at $\mu_p = 1.93~\mu$m. (b) Phase matching function $\Phi(\omega_i,\omega_s)$ for waveguide of length $L=80$ mm and a periodicity of $\Lambda=232~\mu\mathrm{m}$.  (c) The resulting JSA $f(\omega_i,\omega_s)=\alpha(\omega_i+\omega_s)\Phi(\omega_i,\omega_s)$. \emph{The JSA has been plotted as a function of the frequency, however corresponding values for the wavelength have been included.} (d) Schmidt numbers and modes for the JSA (top to bottom): the first 20 Schmidt numbers $b_k$; the first 5 Schmidt modes $\xi_k(\omega_s)$ for the signal state; the first 5 Schmidt modes $\zeta_k(\omega_i)$ for the ilder state, as well as a Gaussian filter function of width $\sigma_f =100$ $\times 10^{9}~\mathrm{s}^{-1}$ (dashed line); the filtered Schmidt modes $T(\omega_i)\zeta_k(\omega_i)$ for the idler state; the othogonalised idler modes $\phi_j(\omega_i)$; the diagonalised signal modes $\tau_m(\omega_s)$}
  \label{fig:JSA_ext2}
\end{figure}

In order to meet the extended phase matching conditions for an asymmetric separable JSA, we again analyse a type II PP-KTP waveguide, of length $L=80$ mm and a periodicity of $\Lambda= 232~\mu\mathrm{m}$, pumped with a $1.93~\mu$m laser with a 3nm FWHM ($\sigma_p=0.64$$\times 10^{12}~\mathrm{s}^{-1}$) which down converts to $3.85~\mu$m in the signal and idler modes. We note that single-photon detection is not particularly practical at this wavelength, however for consistency, we have chosen to use a PP-KTP waveguide throughout this paper. The same JSA can be achieved in different systems, at more practical wavelengths. See, for example, Mosley \emph{et al.} \cite{Mosley2008}.  Unless stated otherwise, the results in this section were obtained using an $800\times 800$ grid, ranging over $8$$\times 10^{12}~\mathrm{s}^{-1}$, centered around $\omega_i=\omega_s=\mu$.

 \begin{figure}[t!]
 \begin{center}
  \vspace{-.2cm}
 \includegraphics[width=7cm]{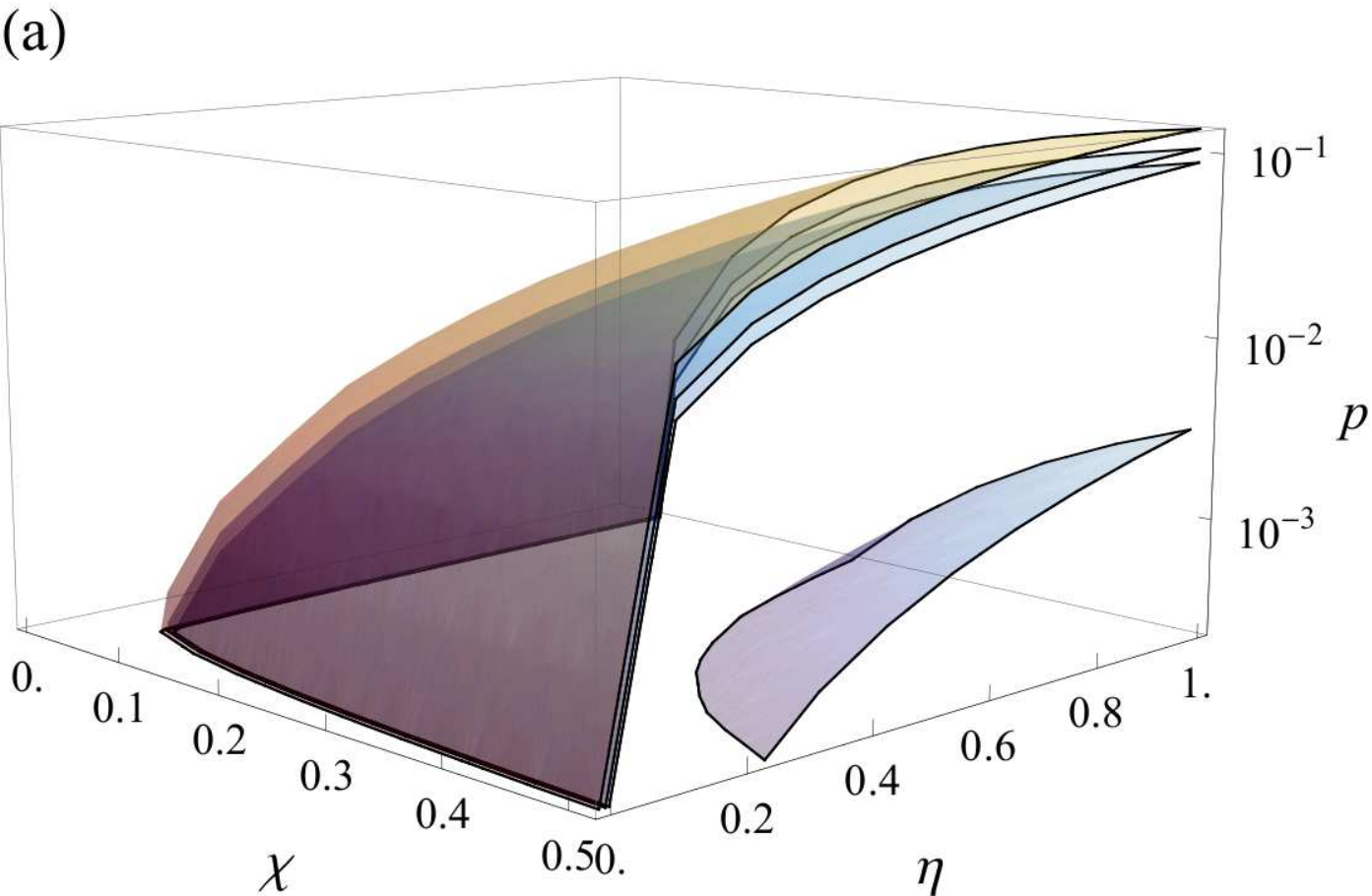} ~~~     \includegraphics[width=7cm]{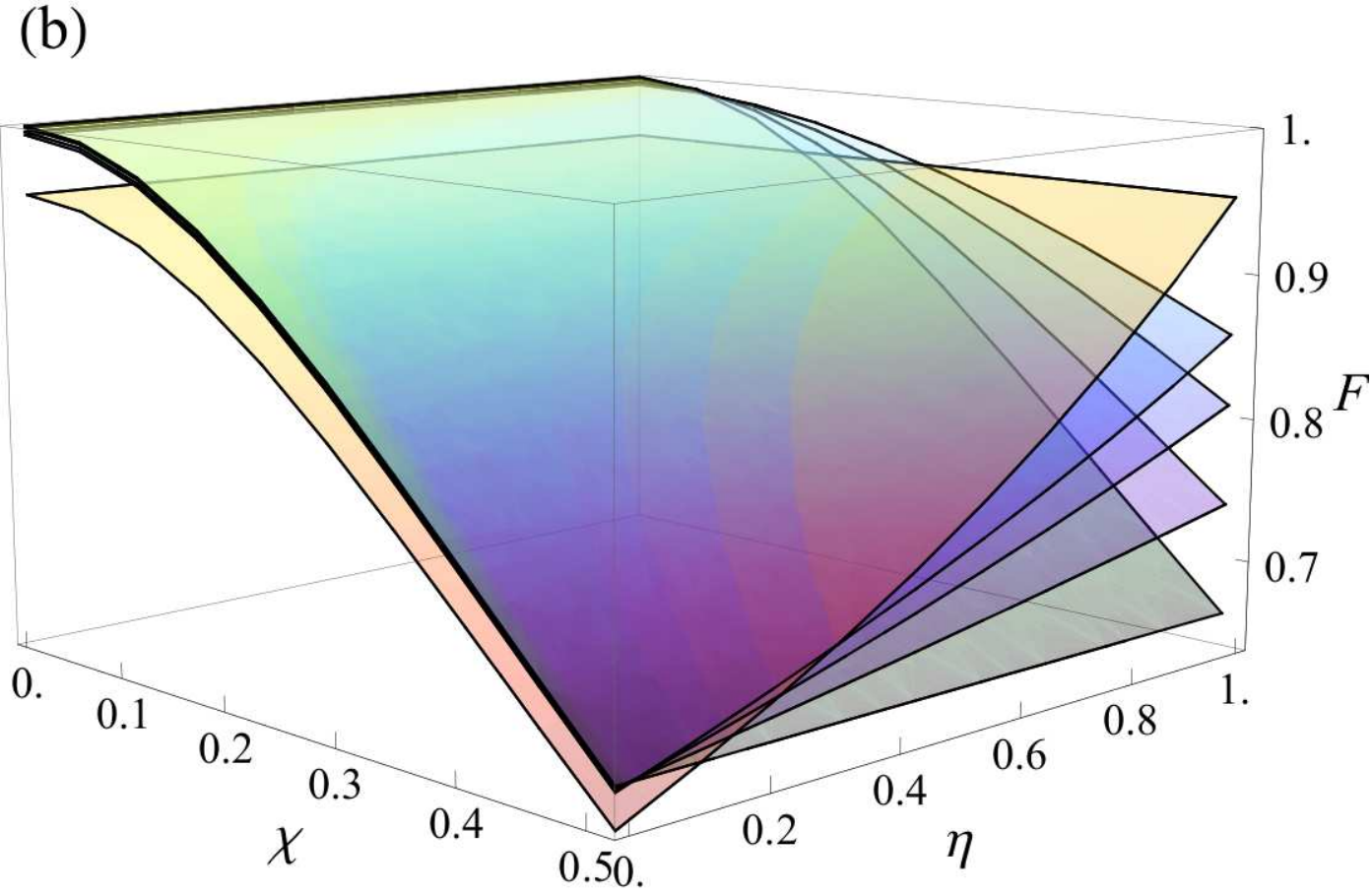}\\
         \includegraphics[width=7cm]{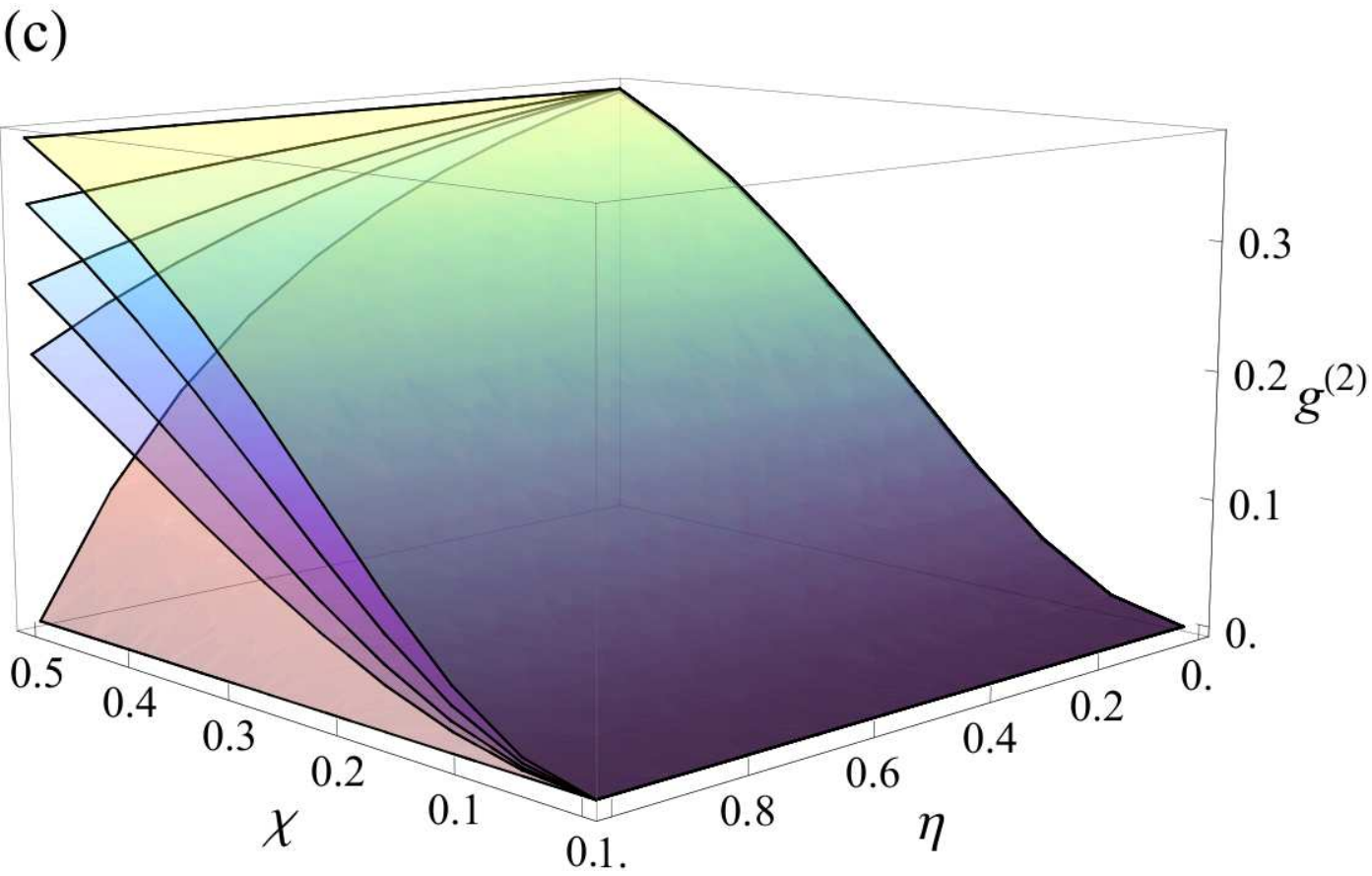}  ~~~ \includegraphics[width=7cm]{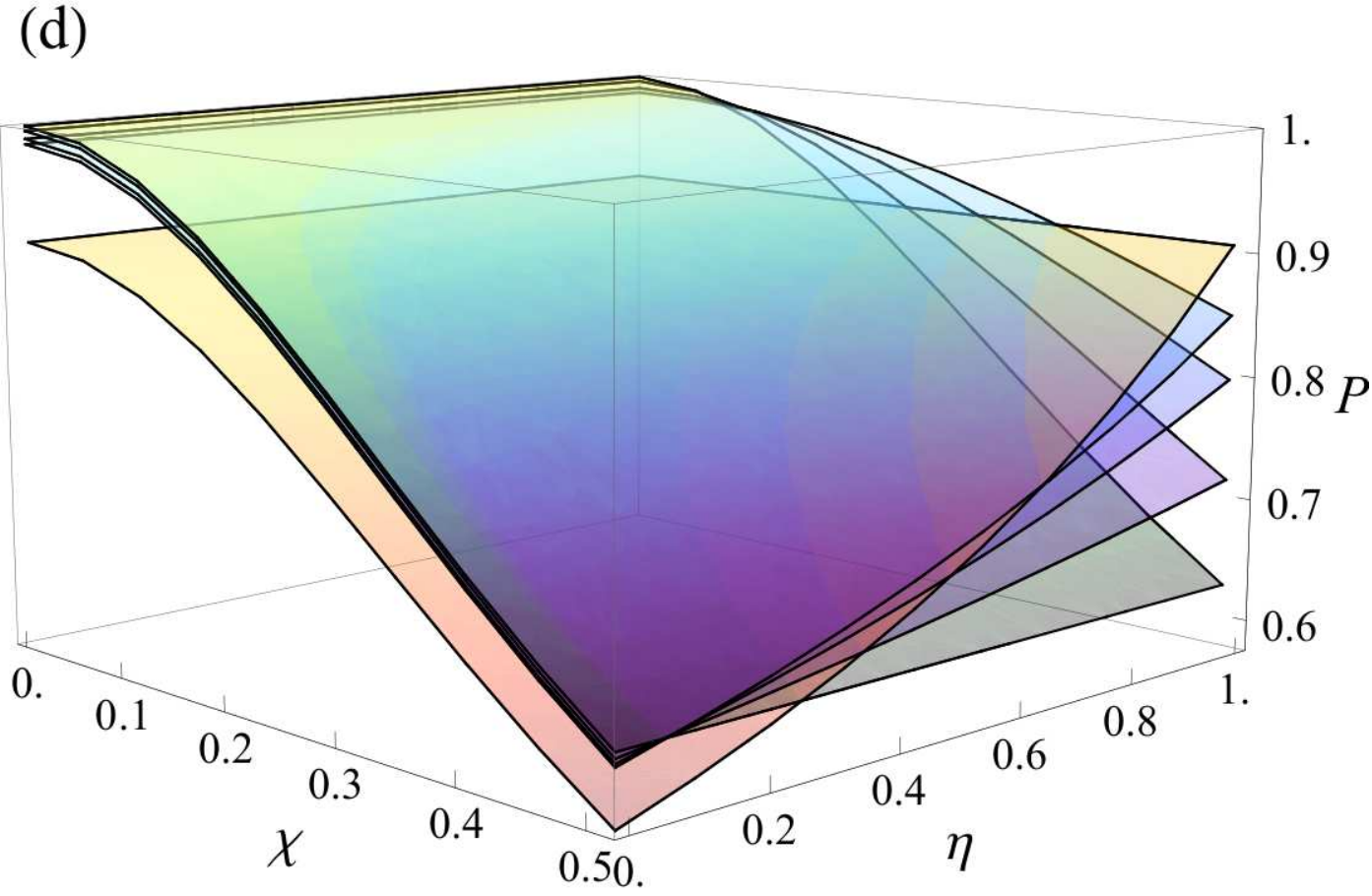}
   \vspace{-.5cm}
\end{center}
  \caption{(a) The probability of detecting a single photon in the idler mode for (top to bottom): no filter; $\sigma_f =150$$\times 10^{9}~\mathrm{s}^{-1}$; $\sigma_f =100$$\times 10^{9}~\mathrm{s}^{-1}$; $\sigma_f =50$$\times 10^{9}~\mathrm{s}^{-1}$. \emph{Note that the probability is plotted on a log scale. }(b)  The fidelity of the signal state with an ideal Fock state for (top to bottom at $\chi=0.5$ and $\eta=1$): no filter; $\sigma_f =150$$\times 10^{9}~\mathrm{s}^{-1}$; $\sigma_f =100$$\times 10^{9}~\mathrm{s}^{-1}$; $\sigma_f =50$$\times 10^{9}~\mathrm{s}^{-1}$; $\sigma_f =0$. (c) The $g^{(2)}$ of the signal state for (top to bottom): $\sigma_f =0$; $\sigma_f =50$$\times 10^{9}~\mathrm{s}^{-1}$; $\sigma_f =100$$\times 10^{9}~\mathrm{s}^{-1}$; $\sigma_f =150$$\times 10^{9}~\mathrm{s}^{-1}$; no filter. \emph{Note the change in axis orientation.} (d) The purity of the signal state for (top to bottom at $\chi=0.5$ and $\eta=1$): no filter; $\sigma_f =150$$\times 10^{9}~\mathrm{s}^{-1}$; $\sigma_f =100$$\times 10^{9}~\mathrm{s}^{-1}$; $\sigma_f =50$$\times 10^{9}~\mathrm{s}^{-1}$; $\sigma_f =0$. } 
  \label{fig:probfid1_ext2} 
\end{figure}

Figure \ref{fig:JSA_ext2} shows the JSA and the corresponding Schmidt values and modes. Notice in the Schmidt decomposition that the first mode is even more dominant than in the symmetric case. The entropy of entanglement for this JSA is $E=0.37$. 

Setting $k_p'=k_s'$ generates a vertical phase matching function. As long as the waveguide is sufficiently long, and therefore, the width of the phase matching function sufficiently thin, and the pump is sufficiently wide, the result will be a vertical, almost elliptical and very thin JSA. 

As in the previous section, we present results for the probability, purity, $g^{(2)}$ and fidelity, for the heralding of one and two photon Fock states. Although it is possible to achieve purities arbitrarily close to unity by increasing the length of the waveguide, the vertical orientation of the JSA places it in a unique position to take advantage of spectral filtering. We will once again compare results for: an unfiltered idler state; an idler state filtered with a Gaussian filter $T(\omega_i)=\exp(-(\mu_f-\omega_i)/2\sigma_f^2)$, of various widths $\sigma_f$ and centered at the central idler frequency; as well as the limiting case where $T(\omega_i)=\delta(\omega_i-\mu_f)$.

\subsubsection{Generating single-photon Fock states}

Figure \ref{fig:probfid1_ext2} (a) shows the probability of detecting a single photon in the signal mode. Since most of the photons will have spectral distributions within the filter width, we do not see a very big drop in the probability with filtering, until the filter is so narrow that it cuts into the central lobe. 

Figure \ref{fig:probfid1_ext2} (b) shows the fidelity. We will distinguish between these surfaces by referring to their values at $\chi=0.5$ and $\eta=1$. Again, filtering the idler mode increases the fidelity of the signal state with a single photon in the low-$\chi$ and low-$\eta$ regimes.

Figure \ref{fig:probfid1_ext2} (c) shows the $g^{(2)}$ and figure \ref{fig:probfid1_ext2} (d) shows the purity of the state in the signal mode.  Again, we will distinguish between these surfaces by referring to their values at the point $\chi=0.5$ and $\eta=1$. Results for the filtered case were computed using a $600\times 600$ grid, ranging over $8$$\times 10^{12}~\mathrm{s}^{-1}$, centered around $\omega_i=\omega_s=\mu$, and truncating $b_k$ with values below $10^{-2}$.

Figure \ref{fig:ChiProb}~(c) shows the probability of success, and required nonlinearity, for a number of filter widths, in order to achieve a fidelity of $F=0.95$, using a heralding detector with efficiency $\eta=0.5$. Note that, as with the symmetric case, while the probability of success is greatly enhanced by using a source engineered state, the required pump powers are nearly the same. Also notice the ``flat'' region, where the probability does not change much, between $\sigma_f=0.3\times10^{12}~\mathrm{s}^{-1}$ and  $0.35\times10^{12}~\mathrm{s}^{-1}$. This corresponds to the ``dark'' region between the lobes on the JSA. Over this region, we do not expect much change in the flux.

\newpage

\subsection{Generating 2-photon Fock states}

\begin{figure}[h!]
  \begin{center}
 \includegraphics[width=7cm]{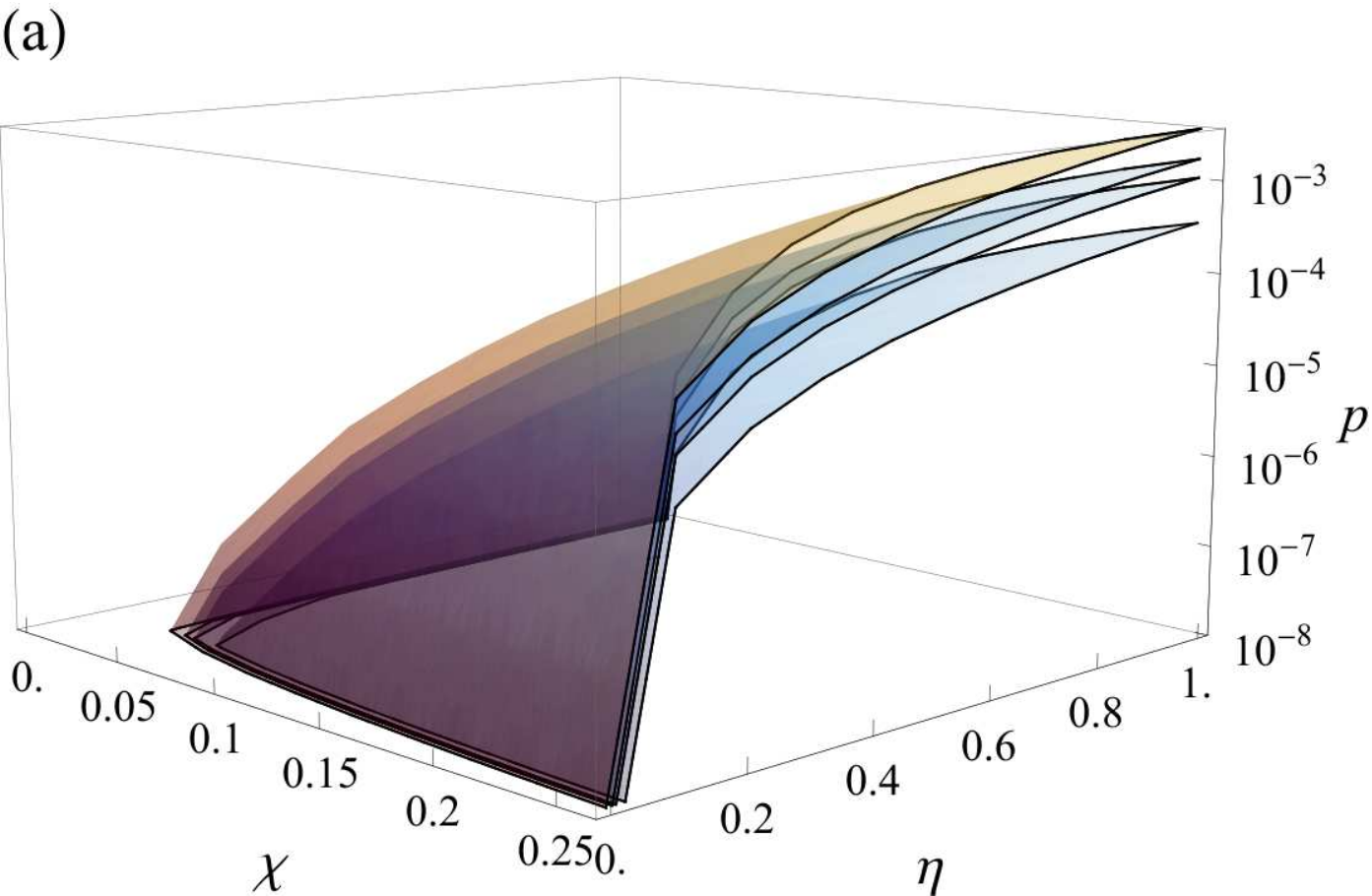} ~~~     \includegraphics[width=7cm]{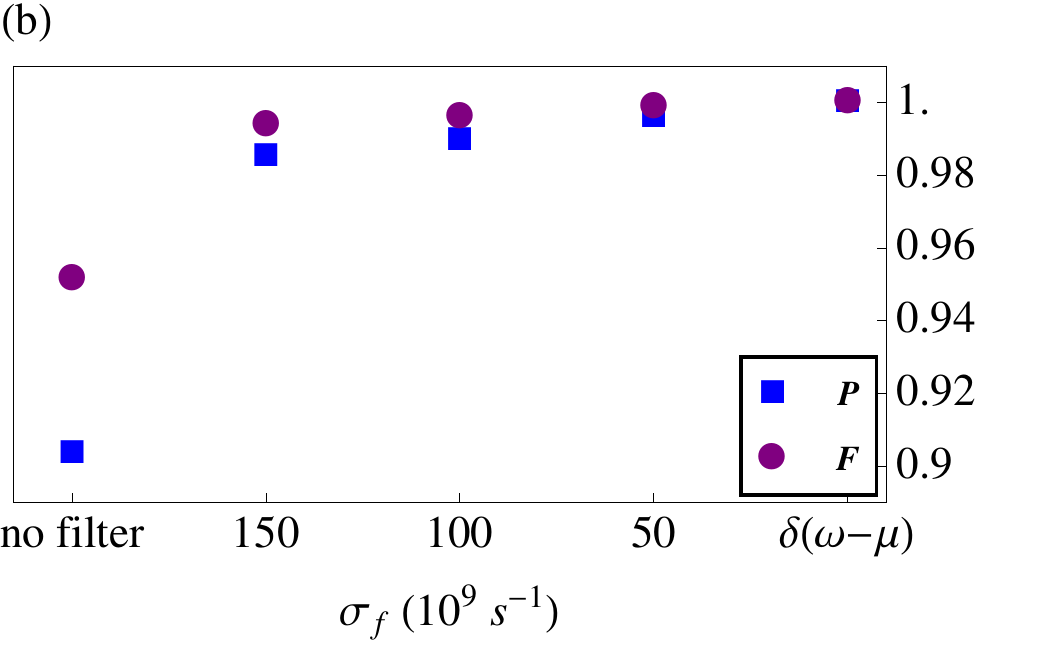}
   \vspace{-.5cm}
  \end{center}
    \caption{(a) The probability of detecting two photons in the idler mode for (top to bottom): no filter; $\sigma_f =150$$\times 10^{9}~\mathrm{s}^{-1}$; $\sigma_f =100$$\times 10^{9}~\mathrm{s}^{-1}$; $\sigma_f =50$$\times 10^{9}~\mathrm{s}^{-1}$. \emph{Note that the probability is plotted on a log scale. }(b)  The purity and fidelity of the signal state with an ideal Fock state. } 
  \label{fig:prob2ext2}
\end{figure}

Figure \ref{fig:prob2ext2} (a), represents the probability of detecting two photons in the idler mode. The corresponding fidelities and purities have been shown in figure \ref{fig:prob2ext} (b). They do not vary as a function of $\chi$ and $\eta$.

\section{Discussion}\label{sec:conc}

We have calculated the spectrally entangled output state of a parametric down converter to second order in photon number, with the goal of generating heralded one- and two-photon Fock states in one spatial mode (signal), conditional on the detection of one or two photons in the other spatial mode (idler). We have presented analytical expressions for the heralded state after the idler mode is spectrally filtered using a Gaussian filter and detected with an inefficient detector. The heralded signal state was then characterised by its $g^{(2)}$ and purity. In addition, we calculated the fidelity of the heralded state with the desired ideal Fock state.

As a physical example, we modeled a type II PP-KTP waveguide, pumped by lasers at wavelengths of $400$ nm, $788$ nm and $1.93~\mu$m. We found that in the first example, where no effort was made to perform any extended phase matching conditions, the results were states with very low purity. After strong spectral filtering, Fock states with arbitrarily high purity could be achieved, however at very low probabilities of success. To achieve a fidelity of $F=0.95$ for a single-photon state, using a heralding detector with efficiency $\eta=0.5$, the probability of success would be on the order of $10^{-4}$. 

The latter two examples, where extended phase matching conditions were fulfilled, resulted in much higher purity states, however, some additional filtering was still required to achieve very high purity states. Both the symmetric and asymmetric examples were able to achieve a fidelity, with a single-photon Fock state, of $F=0.95$, using a heralding detector with efficiency $\eta=0.5$, with probabilities of success on the order of $10^{-2}$. High purity two-photon Fock states were also possible. While results were comparable for the symmetric and asymmetric examples, this was due to our choice of physical parameters. The asymmetric case would be able to achieve higher fidelities, with no filtering, by choosing a longer waveguide. 

\section{Acknowledgments}

AMB acknowledges useful discussions with Alessandro Fedrizzi, Andreas Christ, Andreas Eckstein and Kaisa Laiho. We acknowledge the financial support of the Future and Emerging Technologies (FET) programme within the Seventh Framework for Research of the European Commission, under the FET-Open grant agreement CORNER no. FP7-ICT-213681.

\bibliographystyle{unsrt.bst}

\end{document}